\documentclass[11pt,a4paper]{article}
\usepackage{jcappub}
\usepackage{bm}
\usepackage{amsmath}
\usepackage{xcolor}
\def\dd{\mathrm{d}}

\def\mcP{\mathcal{P}}

\def\mcK{\mathcal{K}}

\def\bk{\mathbf{k}}

\title{Quantum Entanglement in Multi-field Inflation}

\author[a]{Nadia Bolis,}
\author[b]{Tomohiro Fujita,}
\author[c]{Shuntaro Mizuno}
\author[c,d]{and Shinji Mukohyama}

\affiliation[a]{%
Central European Institute for Cosmology and Fundamental Physics, Institute of
Physics, Czech Academy of Sciences, Na Slovance 1999/2 Prague, Czech Republic.}%

\affiliation[b]{%
Department of Physics, Kyoto University, Kyoto 606-8502, Japan
}%

\affiliation[c]{%
Center for Gravitational Physics, Yukawa Institute for Theoretical Physics, Kyoto University, Kyoto 606-8502, Japan
}%

\affiliation[d]{%
 Kavli Institute for the Physics and Mathematics of the Universe (WPI),
The University of Tokyo Institutes for Advanced Study,
The University of Tokyo, Kashiwa, Chiba 277-8583, Japan
}%

\emailAdd{bolis at fzu.cz}
\emailAdd{t.fujita at tap.scphys.kyoto-u.ac.jp}
\emailAdd{shuntaro.mizuno at yukawa.kyoto-u.ac.jp}
\emailAdd{shinji.mukohyama at yukawa.kyoto-u.ac.jp}

\abstract{We study the emergence of quantum entanglement in multi-field inflation. In this scenario, the perturbations of one field contribute to the observable curvature perturbation, while multi-field dynamics with the other fields affect the curvature perturbation through particle production and entanglement. We develop a general formalism which defines the quantum entanglement between the perturbations of the multiple fields both in the Heisenberg and Schr\"odinger pictures, and show that entanglement between different fields can arise dynamically in the context of multi-field inflationary scenarios. We also present a simple model in which a sudden change in the  kinetic matrix of the scalar fields generates entanglement and an oscillatory feature appears in the power spectrum of the inflaton perturbation.
}

\begin{document}

%********************** PREPRINT NUMBERS ****************************%
\begin{flushright}%
 YITP-18-48, IPMU18-0091
\end{flushright}

\maketitle
%\tableofcontents

%************************************************************************************%
%
%
%
%====================================================================================%
\section{Introduction }
%====================================================================================%

Cosmic inflation is widely believed to be the most plausible scenario to explain the origin of temperature fluctuations of the cosmic microwave background (CMB)  and large scale structure (LSS) of the Universe \cite{Guth:1982ec,Hawking:1982cz,Starobinsky:1982ee,Bardeen:1983qw} (see e.g. \cite{Kodama:1985bj,Mukhanov:1990me} for reviews). The fact that the primordial curvature perturbations are almost scale-invariant and Gaussian is strongly supported by the recent Planck observations \cite{Ade:2015xua,Ade:2015ava}. These observations are consistent with the predictions of the simplest single-field inflation models, where the inflaton has a canonical kinetic term and a sufficiently flat potential that allows it to roll slowly during inflation, and couples minimally to gravity. Regardless of these phenomenological successes, however, it is still nontrivial to embed single-field slow-roll inflation into a more fundamental theory such as string theory. To search for new physics in inflation, and a possible connection to a more fundamental theory, one or more of the following conditions can be relaxed: single field, slow-roll or canonical kinetic term (see \cite{Baumann:2014nda}, for a review)

Contrary to the above approach, where the effect of new physics is encoded in some terms in the action, it is also possible that new physics can change the initial quantum state of the inflaton perturbations. Here, by an initial quantum state we mean a quantum state at the time when the system enters the regime of validity of the particular low-energy effective field theory under consideration. In the standard scenario, the primordial power spectrum is computed assuming an initial vacuum state for curved spacetime, the so-called Bunch-Davies (BD) vacuum \cite{Bunch:1978yq}. The behavior of the mode functions of the inflaton perturbations in the BD vacuum reduce to flat spacetime vacuum mode functions in the short distance limit. However, it is certainly conceivable that short distance physics can give rise to deviations from the Bunch-Davies vacuum \cite{Martin:2000xs,Easther:2001fi,Danielsson:2002kx,Easther:2002xe,Danielsson:2002qh,Kaloper:2002uj, Easther:2001fz,Kaloper:2003nv,Ashoorioon:2004vm,Collins:2005nu,Collins:2006bg,Carney:2011hz}, whose effect can be described in terms of particle production. Such particle production is clearly understood by Bogoliubov transformations in the low energy effective theory. Although the choice of initial state is often discussed in the context of trans-Planckian effects (see \cite{Brandenberger:2012aj, Martin:2000xs, Collins:2005nu}, for a review), the issue has more general applicability, for example, a discussion of how to capture the initial state effects in terms of a boundary effective field theory \cite{Schalm:2004qk,Greene:2004np,Schalm:2004xg}.

Furthermore, in fundamental theories like supergravity or string theory, scalar fields are ubiquitous and in some cases, some fields can affect each other through entanglement, even if they are decoupled at the level of the low energy effective action.  Likewise, in the context of the string theory landscape, it was shown that quantum entanglement can exist between two causally disconnected regions in de Sitter space  \cite{Maldacena:2012xp}. Modifications of the power spectrum of CMB generated by such scenarios with quantum entanglement between two causally disconnected universes was studied in \cite{Kanno:2014ifa, Kanno:2015lja, Dimitrakopoulos:2015yva,Kanno:2015lja} (for recent work extending this setup, see \cite{Albrecht:2018prr}).

 Apart from  the multiverse scenario, it is quite interesting to consider the possibility that the inflaton and some other scalar field are entangled initially. In \cite{Albrecht:2014aga}, one of us investigated the cosmological consequences of an entangled initial state between the inflaton and another spectator field during inflation (see for related works \cite{Collins:2016ahj,Bolis:2016vas, Rostami:2017ktl}). In this work, it was shown that an entangled initial state for a toy model consisting of non-interacting, minimally coupled scalar fields in a fixed de Sitter background will produce small oscillations in the power spectrum of the inflaton perturbations. The entangled initial state ansatz in \cite{Albrecht:2014aga} was set to be Gaussian and the strength of the entanglement is parametrized by one parameter. In particular, the quantum state of the whole system includes entanglement between the two fields, while the Hamiltonian does not have any interaction terms between them. In this first study, the entangled initial state is used as a tool to phenomenologically test for small deviations from a BD initial state. However, the dynamical origin of the entanglement encoded in the initial state and possible connections to fundamental theories were not explored explicitly. Furthermore, whether the oscillations are a generic feature of an entangled initial state or just come from a specific class of initial states was also not discussed \cite{Kanno:2015ewa}.

The aim of this paper is to address these concerns by showing that entanglement between different fields can arise dynamically in the context of multi-field inflation scenarios. Our analysis is restricted to consider perturbations of scalar fields in a fixed de Sitter background. 
We will formulate the emergence of entanglement between
the perturbations of two scalar fields on a general basis,
and we will illustrate this phenomena by studying a concrete example of this type of model.

The rest of this paper is organized as follows. In section~\ref{sec_general argument}, we derive within the Heisenberg picture the general condition under which a state, from the view point of a late time observer, is entangled. For this discussion, we consider a scenario in which the vacuum naturally defined at sufficiently late times becomes different from that at sufficiently early times, as a result of the time evolution of the kinetic and mass matrices in multi-field inflation. In section~\ref{sec_schrodinger}, we make the connection to the state generated in the  Schr\"odinger picture and see how our initial-vacuum state corresponds to the initial state considered in  \cite{Albrecht:2014aga}. Then, in section~\ref{sec_concrete model} we  consider a concrete example that produces an entangled state by the effect of kinetic mixing  and confirm that the oscillations in the power spectrum of the inflaton perturbations are also obtained in this model. Section~\ref{sec_conclusion} is devoted  to conclusion and discussion.

%====================================================================================%
\section{Entanglement in Heisenberg Picture\label{sec_general argument}}
%====================================================================================%

%====================================================================================%
\subsection{Entangled state at late-time
%Late-time setup
}\label{subsec_latetimesetup}
%====================================================================================%

In this subsection, we describe the quantum state of interest, while in the following subsections we will discuss a physical mechanism that can give rise to such a state. 
At late times, say $t\geq t_0$, in any globally hyperbolic spacetime region, let us consider two scalar fields $\phi$ and $\sigma$ and suppose that interactions between them can be ignored. Let us further assume that self-interactions of $\phi$ and $\sigma$ are weak and can be ignored (or can be treated perturbatively). One can then construct the Hilbert space of the system as 
\begin{equation}
 \mathcal{F} = \mathcal{F}_{\phi}\bar{\otimes}\mathcal{F}_{\sigma}\,,
\end{equation}
where $\bar{\otimes}$ is a tensor product followed by a suitable completion and $\mathcal{F}_{\phi}$ and $\mathcal{F}_{\sigma}$ are the Fock spaces of $\phi$ and $\sigma$, respectively, defined as
\begin{eqnarray}
 \mathcal{F}_{\phi} & =  &
        \bm{C} \oplus \mathcal{H}_{\phi} \oplus 
        \left(\mathcal{H}_{\phi}\otimes\mathcal{H}_{\phi}\right)_{\rm sym} 
        \oplus \cdots\,,\nonumber\\
 \mathcal{F}_{\sigma} & =  &
        \bm{C} \oplus \mathcal{H}_{\sigma} \oplus 
        \left(\mathcal{H}_{\sigma}\otimes\mathcal{H}_{\sigma}\right)_{\rm sym} 
        \oplus \cdots\,. 
\end{eqnarray} 
Here, $\mathcal{H}_{\phi}$ and $\mathcal{H}_{\sigma}$ are the Hilbert spaces of positive-frequency mode functions for $\phi$ and $\sigma$, respectively, and $(\cdots)_{\rm sym}$ denotes the symmetrization ($(\xi\otimes\eta )_{\rm sym}= (1/2)(\xi\otimes\eta +\eta\otimes\xi )$, and so on.). Physically, $\bm{C}$ denotes the vacuum state, $\mathcal{H}_{\phi,\sigma}$ one particle states, $\left(\mathcal{H}_{\phi,\sigma}\otimes\mathcal{H}_{\phi,\sigma}\right)_{\rm sym}$ two particle states, and so on.

We suppose that all our observables are operators on $ \mathcal{F}_{\phi}$ so that $\sigma$ is unobservable. In this case, for a quantum state $|\psi\rangle$ ($\in \mathcal{F}$) the expectation value of an observable $\mathcal{O}$ is 
\begin{equation}
 \langle \psi | \mathcal{O} |\psi\rangle = {\bf Tr}_{\phi} [\mathcal{O}\rho_{\phi}]\,,
\end{equation}
where
\begin{equation}
 \rho_{\phi} \equiv {\bf Tr}_{\sigma}|\psi\rangle \langle\psi |\,,
\end{equation}
is the reduced density matrix, and ${\bf Tr}_{\phi}$ and ${\bf Tr}_{\sigma}$ are the trace operations in $\mathcal{F}_{\phi}$ and $\mathcal{F}_{\sigma}$, respectively. In general, the reduced density matrix $\rho_{\phi}$ represents a mixed state due to quantum entanglement between $\mathcal{F}_{\phi}$ and $\mathcal{F}_{\sigma}$ unless the quantum state $|\psi\rangle$ of the total system is a direct product of a state in $\mathcal{F}_{\phi}$ and a state in $\mathcal{F}_{\sigma}$.

Next let us consider possible ``initial'' states of the two-field system at $t=t_0$. Since by the assumption the fields $\phi$ and $\sigma$ for $t\geq t_0$ do not interact with each other, it may be natural to consider a direct product of a state $|\psi_{\phi}\rangle_{\phi}$ in $\mathcal{F}_{\phi}$ and a state $|\psi_{\sigma}\rangle_{\sigma}$ in $\mathcal{F}_{\sigma}$. 
\begin{equation}
 |\psi_{\phi}\rangle_{\phi}\otimes |\psi_{\sigma}\rangle_{\sigma}\,, \qquad
  (\mbox{direct product state})\,. \label{eqn:directproduct}
\end{equation}
This class of states does not contain entanglement between $\mathcal{F}_{\phi}$ and $\mathcal{F}_{\sigma}$ and the reduced density matrix represents a pure state as $\rho_{\phi}=|\psi_{\phi}\rangle_{\phi}{}_{\phi}\langle\psi_{\phi}|$. Starting with the direct product state (\ref{eqn:directproduct}), 
%if one likes, one can add 
one can consider a Gaussian-type entanglement between $\mathcal{F}_{\phi}$ and $\mathcal{F}_{\sigma}$ as
\begin{equation}
 |\psi,\mathcal{C}\rangle = \mathcal{N}\exp\left[\frac{1}{2}\sum_{m,n}\mathcal{C}^{mn}\hat{a}_m^{\dagger}\hat{b}_n^{\dagger}\right]\, |\psi_{\phi}\rangle_{\phi}\otimes |\psi_{\sigma}\rangle_{\sigma}\,, \qquad
  (\mbox{entangled state}), \label{eqn:entangled}
\end{equation}
where $\{\hat{a}_m^{\dagger}\}$ and $\{\hat{b}_n^{\dagger}\}$ are sets of creation operators associated with $\mathcal{H}_{\phi}$ and $\mathcal{H}_{\sigma}$, respectively, $\mathcal{C}^{mn}$ is a matrix characterizing the entanglement to be added and $\mathcal{N}$ is a normalization constant. As we shall see in the rest of this section, the multi-field dynamics in the early epoch ($t<t_0$) generically ends up with this type of quantum state at $t=t_0$. As we shall see in Sec.~\ref{sec_schrodinger}, a quantum state considered in \cite{Albrecht:2014aga} is also of this type.

%====================================================================================%
\subsection{Late-time and early-time setup }\label{subsec_setup} 
%====================================================================================%

In the previous subsection we introduced %a late-time setup
the entangled state in eq.~\eqref{eqn:entangled} for $t>t_0$, where the two scalar fields $\phi$ and $\sigma$ do not have direct interactions at the level of the Lagrangian but the initial state includes entanglement between these fields. In the present subsection, in order to discuss a possible explanation for the physical origin of such entanglement, we shall consider a setup for an earlier time, whose final state corresponds to the initial state for the late-time setup of the previous subsection. We shall show that the entanglement contained in the initial state of the late-time setup naturally emerges from the early-time multi-field dynamics, starting with the standard Bunch Davies vacuum state.

We consider the following simple action for the two scalar fields $\phi^I = \{\phi,\sigma\}$,
\begin{equation}
 S=-\frac{1}{2} \int \sqrt{-g} \dd^4x\,
  \left[ G_{IJ}(\varphi) g^{\mu\nu}\partial_\mu \phi^I  \partial_\nu \phi^J
   + M_{IJ}(\varphi) \phi^I\phi^J \right]\,, \label{first action}
\end{equation}
in a flat Friedmann-Lema\^{i}tre-Robertson-Walker (FLRW) background,
\begin{equation}
 g_{\mu\nu} dx^{\mu}dx^{\nu} = -dt^2 + a(t)^2(dx^2+dy^2+dz^2)\,,
\end{equation}
where $G_{IJ}(\varphi)$ is a field space metric, $M_{IJ}(\varphi)$ is a squared mass matrix and $\varphi$ is another field. We shall not specify the action for $\varphi$ but assume that it is homogeneous and time-dependent so that 
$G_{IJ} (\varphi (t))$ and $M_{IJ}  (\varphi (t))$ are $2\times2$ real symmetric matrices that evolve in time. In what follows, with the possibility that the time dependence of 
$G_{IJ}$ and $M_{IJ}$ is realized through the dynamics of the field $\varphi$ in mind, 
we phenomenologically treat them as $G_{IJ} (t)$ and $M_{IJ}(t)$. We further assume that the backreaction of $\phi^I = \{\phi,\sigma\}$ on the background is small enough.

In accordance with the previous subsection, we assume that $G_{IJ}$ and $M_{IJ}$ are diagonal and constant for $t\geq t_0$. Otherwise, the two fields would interact with each other at late times, either directly or through $\varphi$. After canonically normalizing $\phi^I = \{\phi,\sigma\}$, we thus have
\begin{equation}
G_{IJ}(t) = \bm{1}, \qquad M_{IJ} (t) = {\rm diag}(m_\phi^2, m_\sigma^2)\,.
 \qquad ({\rm late\ time, i.e.}\  t\geq t_0)
\label{GM diagonalized late}
\end{equation}

On the other hand, at sufficiently early times, %we assume that 
$G_{IJ} (t)$ %itself is not a unit matrix but
is, in general, diagonalized and normalized by another set of fields $\Phi^I=\{\Phi,S\}$; \footnote{This order is chosen
so that $\Phi = \phi$ and $S = \sigma$ if $G_{IJ}$ has no time evolution at all, that is, if  $G_{IJ} = \bm{1}$.}
\begin{equation}
G_{IJ}(t) \partial_\mu \phi^I \partial^\mu \phi^J \to 
\delta_{IJ} \partial_\mu \Phi^I \partial^\mu \Phi^J\,,
\qquad ({\rm early\ time})
\label{GM diagonalized early}
\end{equation}
where $\Phi^I$ and $\phi^I$ are related to each other by a $2\times2$ real matrix $\mcK_{IJ}$ as 
\begin{equation}
\phi^I = \mcK_{IJ} \Phi^J\,.
\label{relation two sets of asymptotic fields}
\end{equation}

In this setup for simplicity, we consider the perturbations of the scalar fields up to quadratic order
in the action
in a fixed de Sitter background,  where the perturbations and the background are decoupled.
%, aside from the Hubble parameter $H\equiv \dot{a}/a$. 
Following ref.~\cite{Albrecht:2014aga}, we take $\phi$ to be the inflaton 
whose fluctuations are related with the observable curvature perturbation, while
$\sigma$ is the non-inflaton field whose fluctuations are decoupled from the curvature perturbation
after inflation. In what follows, we investigate 
%seek 
the influence of the changes in $G_{IJ} (t)$ and $M_{IJ} (t)$
during inflation on the perturbations of the inflaton $\phi$, with special emphasis on
how it can induce quantum entanglement with the other spectator field $\sigma$.
%, by constructing two asymptotic vacuum states
%as well as the reduced density matrix for $\phi$.

%====================================================================================%
\subsection{In-vacuum state and out-vacuum state \label{subsec_invacuum_outvacuum}}
%====================================================================================%

Here,  we construct the asymptotic vacuum state at sufficiently early times and late times, in order  to discuss the quantum entanglement between $\phi$ and $\sigma$.

In the far past, %from the discussions in subsection~\ref{subsec_setup}, 
the in-vacuum state is constructed from $\Phi$ and $S$,
since they have canonical kinetic terms.
 Decomposing these fields into the homogeneous (background) and inhomogeneous (perturbation) parts,
\begin{equation}
\Phi(t,\bm x)=\Phi_0(t)+\delta\Phi(t,\bm x),\qquad
S(t,\bm x)=S_0(t)+\delta S(t,\bm x),
\end{equation}
and introducing canonical %Mukhanov-Sasaki like 
variables
\begin{equation}
u_\Phi \equiv a \delta \Phi, \qquad
u_S \equiv a \delta S,
\end{equation}
one can rewrite the Lagrangian density at early times for the perturbations as
\begin{equation}
\mathcal{L}(\eta,\bm{x})=\frac{1}{2}%\int \dd \eta\, \dd^3x\, 
\left[ 
{u_\Phi'}^2- (\partial_i u_\Phi)^2+{u_S'}^2- (\partial_i u_S)^2-a^2 (\tilde{M}_{IJ}- 2 H^2 \delta_{IJ} ) u_\Phi^I  u_\Phi^J
\right],\quad ({\rm early \ time})
\label{past action}
\end{equation}
where we converted to conformal time $\eta$ which is related to the scale factor by $a= -1/(H \eta)$
and the prime denotes $\partial_\eta$. The new mass matrix in terms of the original one is 
$\tilde{M}_{IJ} = (\mcK^T M \mcK)_{IJ}$ and $u_\Phi ^I=\{u_\Phi, u_S\}$. 
%The perturbations are redefined as $\Phi\propto a \delta\phi$ and $S\propto a\delta\sigma$ such that the kinetic term is diagonalized, and they have a new mass matrix $\tilde{M}_{IJ}(t)$ whose time dependence is weak.
We can then quantize these fields with the standard procedure,
\begin{align}
\hat{u}_{\Phi}(\eta,\bm{x}) &=\int \frac{\dd^3 k}{(2\pi)^3}e^{i\bm{k}\cdot \bm{x}}
\hat{u}_\Phi (\eta, {\bm k}),\qquad  \hat{u}_\Phi (\eta, {\bm k}) = u_{\Phi k}(\eta)   \hat{a}_{\bm k} ^{(\Phi)}  +
 u_{\Phi k}^* (\eta) ( \hat{a}_{- \bm k} ^{(\Phi)} )^\dag,\label{uPhi expansion}
\\
\hat{u}_{S}(\eta,\bm{x}) &=\int \frac{\dd^3 k}{(2\pi)^3}e^{i\bm{k}\cdot \bm{x}}
\hat{u}_S (\eta, {\bm k}),\qquad  \hat{u}_S (\eta, {\bm k}) = u_{S k}(\eta)   \hat{b}_{\bm k} ^{(S)}  +
 u_{S k}^* (\eta) ( \hat{b}_{- \bm k} ^{(S)} )^\dag,\label{uS expansion}
\end{align}
where $ \hat{a}_{\bm k} ^{(\Phi)}, (\hat{a}_{ \bm k} ^{(\Phi)} )^\dag,  \hat{b}_{\bm k}  ^{(S)}, 
 (\hat{b}_{\bm k}  ^{(S)})^\dag$ are creation/annihilation operators.
The evolution equations for the mode functions  $u_{\Phi k}^I =\{u_{\Phi k}, u_{S k}\}$ 
%are derived by setting the variation of the action (\ref{past action}) with respect to $u_\Phi ^I = 0$ and 
on sufficiently small scales become
%we have
%
\begin{equation}
{u_{\Phi k} ^I} '' + k^2 u_{\Phi k} ^I \simeq 0.
\label{mode equation early small}
\end{equation}
These mode functions are normalized so that they satisfy the Wronskian condition
\begin{equation}
u_{\Phi k} ^I {u_{\Phi k} ^{I *}} ' - {u_{\Phi k} ^{I}} '  u_{\Phi k} ^{I *} = i, 
\label{mode equation early small}
\end{equation}
which ensures that the creation/annihilation operators satisfy the commutation relations,
\begin{equation}
[ \hat{a}_{\bm k} ^{(\Phi)}, (\hat{a}_{ \bm {k}'} ^{(\Phi)} )^\dag ]
= %(2\pi)^3\delta(\bm{k}-\bm{k}'),\qquad
[ \hat{b}_{\bm k}  ^{(S)}, (\hat{b}_{\bm {k}'}  ^{(S)})^\dag ]= (2\pi)^3\delta(\bm{k}-\bm{k}'),
\qquad
({\rm others})=0.
\label{commurelation early}
\end{equation}
The initial conditions for the  mode functions $u_{\Phi k}(\eta)$ and $u_{S k}(\eta)$ are given, 
to a good approximation,  by the Bunch-Davies vacuum
\begin{equation}
u_{\Phi k} ^{\rm BD} \simeq \frac{e^{-ik\eta}}{\sqrt{2k}}, \qquad
u_{S k} ^{\rm BD} \simeq \frac{e^{-ik\eta}}{\sqrt{2k}},
\end{equation}
% $e^{-ik\eta}/\sqrt{2k}$ 
because their masses (and the mass mixing effect) are negligible deep inside the horizon.\footnote{If the diagonalized mass is larger than $3H/2$, the Bunch-Davies solution should be multiplied by an exponential factor, $\exp[-\frac{\pi}{2}\sqrt{m^2/H^2-9/4}]$, which is otherwise a constant phase. } 
We define the Fock vacuum and construct an in-vacuum state which is annihilated by
the $\hat{a}_{\bm k} ^{(\Phi)}$ and $\hat{b}_{\bm k}  ^{(S)}$ operators,
\begin{equation}
\hat{a}_{\bm k} ^{(\Phi)} |0\rangle_\Phi=0, \qquad
\hat{b}_{\bm k}  ^{(S)}|0\rangle_S=0,  \qquad
|0\rangle_{\rm in}\equiv %|0\rangle_{ab}= 
|0\rangle_\Phi \otimes |0\rangle_S\,,
\label{def_in_vacuum}
\end{equation}
where ${\bm k}$ denotes all relevant modes.
Note that even if the masses are not completely negligible in the equation of motion for 
 $u_{\Phi k}(\eta)$ and $u_{S k}(\eta)$, we can define the initial
state as long as $\tilde{M}_{IJ}$ is diagonal and WKB solutions are available.

%%====================================================================================%
%\subsection{Late time}
%%====================================================================================%

At sufficiently late times,    the out-vacuum state is constructed from $\phi$ and $\sigma$
which now also have canonical kinetic terms and are decoupled.
Although the procedure to construct the out-vacuum state is straightforward 
and merely requires replacing $\Phi \to \phi$ and $S \to \sigma$ in the discussion 
of the in-vacuum state, for completeness,
we summarize its construction.
In terms of %Mukhanov-Sasaki like 
canonical variables corresponding to $\phi$ and $\sigma$,
the Lagrangian density at late times for the perturbations is
\begin{equation}
\mathcal{L}(\eta,\bm{x})=\frac{1}{2}\left[ 
{u_\phi'}^2- (\partial_i u_\phi)^2+{u_\sigma '}^2- (\partial_i u_\sigma)^2-a^2 (M_{IJ}- 2 H^2 \delta_{IJ} ) u_\phi^I  u_\phi^J
\right],\quad ({\rm late\ time})
\label{late time action}
\end{equation}
with $u_\phi ^I=\{u_\phi, u_\sigma\}$.
We can quantize these fields
based on the expansion of the operator $\hat{u}_\phi ^I$,
\begin{align}
\hat{u}_{\phi}(\eta,\bm{x}) &=\int \frac{\dd^3 k}{(2\pi)^3}e^{i\bm{k}\cdot \bm{x}}
\hat{u}_\phi (\eta, {\bm k}),\qquad  \hat{u}_\phi (\eta, {\bm k}) = u_{\phi k}(\eta)   \hat{a}_{\bm k} ^{(\phi)}  +
 u_{\phi k}^* (\eta) ( \hat{a}_{- \bm k} ^{(\phi)} )^\dag,\label{uphi expansion}
\\
\hat{u}_{\sigma}(\eta,\bm{x}) &=\int \frac{\dd^3 k}{(2\pi)^3}e^{i\bm{k}\cdot \bm{x}}
\hat{u}_\sigma (\eta, {\bm k}),\qquad  \hat{u}_\sigma (\eta, {\bm k}) = u_{\sigma k}(\eta)   \hat{b}_{\bm k} ^{(\sigma)}  +
 u_{\phi k}^* (\eta) ( \hat{b}_{- \bm k} ^{(\sigma)} )^\dag,\label{usigma expansion}
\end{align}
with  creation/annihilation operators $\hat{a}_{\bm k} ^{(\phi)}, (\hat{a}_{ \bm k} ^{(\phi)} )^\dag,  
\hat{b}_{\bm k}  ^{(\sigma)}, (\hat{b}_{\bm k}  ^{(\sigma)})^\dag$ satisfying
\begin{equation}
[ \hat{a}_{\bm k} ^{(\phi)}, (\hat{a}_{ \bm {k}'} ^{(\phi)} )^\dag ]= 
%(2\pi)^3\delta(\bm{k}-\bm{k}'),\qquad
[ \hat{b}_{\bm k}  ^{(\sigma)}, (\hat{b}_{\bm {k}'}  ^{(\sigma)})^\dag ]= (2\pi)^3\delta(\bm{k}-\bm{k}'),
\qquad
({\rm others})=0.
\label{commurelation late}
\end{equation}
%
%Once 
The mode functions $u_{\phi k} ^I=\{u_{\phi k}, u_{\sigma k}\} $ are properly normalized  to satisfy
\begin{equation}
u_{\phi k} ^I {u_{\phi k} ^{I *}} ' - {u_{\phi k} ^{I}} '  u_{\phi k} ^{I *} = i.
\label{mode equation late}
\end{equation}
To construct the out-vacuum, we adopt the solutions for the mode functions $u_{\phi k}(\eta)$ and $u_{\sigma k}(\eta)$ 
with the Bunch-Davies initial condition
$u_{\phi k} \simeq u_{\sigma k} \simeq  e^{-ik\eta}/\sqrt{2k}$
 in the case where 
both $G_{IJ}$ and $M_{IJ}$  did not evolve in time but stay as 
eq.~\eqref{GM diagonalized late} for all time,
\begin{align}
&u_{\phi k}(x) =
\sqrt{\frac{\pi x}{4k}}H_{\nu_\phi}^{(1)}(x),
\qquad \nu_\phi\equiv \sqrt{\frac{9}{4}-\frac{m_\phi^2}{H^2}},
\label{uphi_BD}
\\
&u_{\sigma k}(x)=
\sqrt{\frac{\pi x}{4k}}H_{\nu_\sigma}^{(1)}(x),
\qquad \nu_\sigma \equiv\sqrt{\frac{9}{4}-\frac{m_\sigma^2}{H^2}},
\label{usigma_BD}
\end{align}
where $x\equiv -k\eta$ and $H_\nu^{(1)}(x)$ is the Hankel function of the first kind.
It should be noted that these mode functions satisfy the equations of motion only at late times, because the actual Lagrangian is different from eq.~\eqref{late time action} at  earlier times. 
Now we can define the out-vacuum state as the direct product of the Fock vacuum of 
$\hat{a}_{\bm k} ^{(\phi)}$ and $\hat{b}_{\bm k} ^{(\sigma)}$ in the same way as the in-vacuum state,
\begin{equation}
\hat{a}_{\bm k} ^{(\phi)}|0\rangle_\phi=0, \qquad
\hat{b}_{\bm k} ^{(\sigma)}|0\rangle_{\sigma}=0,  \qquad
|0\rangle_{\rm out}\equiv |0\rangle_\phi \otimes |0\rangle_{\sigma}\,,
\label{out vac def}
\end{equation}
where again ${\bm k}$ denotes all relevant modes.

%====================================================================================%
\subsection{Relation between in-vacuum and out-vacuum}
%====================================================================================%

Although the in-vacuum state and out-vacuum state constructed in
subsection~\ref{subsec_invacuum_outvacuum} are of the same form,
in the presence of non-trivial time evolution of $G_{IJ} (t)$ and $M_{IJ}(t)$, the two asymptotic vacuum states
are in general different. 
Since $\phi$ and $\sigma$ are a mixture of $\Phi$ and $S$ due to 
the time evolution of kinetic and mass matrices, the canonical fields at sufficiently late times
$\hat{u}_\phi$ and $\hat{u}_\sigma$ generally inherit all of the creation/annihilation operators of $\hat{u}_\Phi$ and $\hat{u}_S$.
Thus, in general, the annihilation operators for the out-vacuum can be written as  linear combinations of the creation/annihilation operators
for the in-vacuum as
\begin{align}
\hat{a}_{\bm k} ^{(\phi)}  &= \alpha_k \hat{a}_{\bm k} ^{(\Phi)} 
+\beta_k (\hat{a}_{- \bm {k}} ^{(\Phi)} )^\dag 
+\gamma_k \hat{b}_{\bm k} ^{(S)} + \delta_k\, (\hat{b}_{-\bm {k}}  ^{(S)})^\dag\,,
\label{A expression}
\\
\hat{b}_{\bm k} ^{(\sigma)} &= \bar\alpha_k \hat{a}_{\bm k} ^{(\Phi)} +
\bar\beta_k (\hat{a}_{- \bm {k}} ^{(\Phi)} )^\dag   
+\bar\gamma_k \hat{b}_{\bm k} ^{(S)}+\bar \delta_k\, (\hat{b}_{-\bm {k}}  ^{(S)})^\dag\,,
\label{B expression}
\end{align}
where $\alpha_k, \beta_k, \gamma_k,\delta_k, \bar\alpha_k, \bar\beta_k, \bar\gamma_k,$ and $\bar\delta_k$ are constant coefficients. 
%In the following, we omit the subscript $k$ of these coefficients.
These coefficients can be computed once the time evolution of $G_{IJ}$ and $M_{IJ}$ is fixed, as we will see 
in a concrete example in Section.~\ref{sec_concrete model}.
%in the next section. 
Since $\hat{a}_{\bm k} ^{(\phi)}$ and $\hat{b}_{\bm k} ^{(\sigma)}$ satisfy the commutation relations
(\ref{commurelation late}), %
%\begin{equation}
%[\hat{A}_{\bm k}, \hat{A}^\dag_{\bm{ k}'}]= (2\pi)^3\delta(\bm{k}-\bm{k}'),
%\qquad
%[\hat{B}_{\bm k}, \hat{B}^\dag_{\bm{ k}'}]= (2\pi)^3\delta(\bm{k}-\bm{k}'),
%\qquad
%({\rm others})=0,
%\end{equation}
%
the coefficients automatically satisfy
\begin{align}
|\alpha_k|^2-|\beta_k|^2+|\gamma_k|^2-|\delta_k|^2&=1,
\label{cond1}
\\
|\bar \alpha_k|^2-|\bar \beta_k|^2+|\bar \gamma_k|^2-|\bar \delta_k|^2&=1,
\\
\alpha_k\bar{\beta_k}-\beta_k\bar{\alpha_k}+\gamma_k\bar{\delta_k}-\delta_k\bar{\gamma_k}&=0,
\\
\alpha_k\bar{\alpha}_k^*-\beta_k\bar{\beta}_k^*+\gamma_k\bar{\gamma}_k^*-\delta_k\bar{\delta}_k^*&=0.
\label{cond4}
\end{align}
%
%where we suppressed the subscripts, while the all the coefficients have $k$.

Eqs.~\eqref{A expression} and \eqref{B expression} are the generalized  Bogoliubov transformation. 
If the coefficients $\gamma_k$ and $\delta_k$ ($\bar \alpha_k$ and $\bar \beta_k$) vanished, 
the remaining $\alpha_k$ and $\beta_k$ ($\bar \gamma_k$ and $\bar \delta_k$) would be the Bogoliubov coefficients, for each field respectively.
In our case, however, we have these additional coefficients which come from the mixture between the fields. Since the annihilation operator of $\phi$
contains the creation operators of $\Phi$ and $S$, we find $\phi_{\bm k}$ particles
even in the in-vacuum state,
\begin{equation}
{}_{\rm in}\langle0|\hat{N}^{(\phi)}_{\bm k}|0\rangle_{\rm in}
={}_{\rm in}\langle0|(\hat{a}^{(\phi)}_{\bm k})^\dag\hat{a}^{(\phi)}_{\bm k}|0\rangle_{\rm in}=\left(|\beta_k|^2+|\delta_k|^2\right)\delta(\bm 0).
\end{equation}
The time evolution of $G_{IJ}$ and $M_{IJ}$
causes particle production thanks to
the generalization of the Bogoliubov transformation.
Nevertheless, we shall see below that another interesting quantum phenomena also takes place in this system. 

For later convenience, we also express the annihilation operators for the {\it in}-vacuum in terms of  
the linear combination of the creation/annihilation operators
for the {\it out}-vacuum as
\begin{align}
\hat{a}_{\bm k} ^{(\Phi)}  &=\alpha_k ^* \hat{a}_{\bm k} ^{(\phi)}  -\beta_k (\hat{a}_{- \bm {k}} ^{(\phi)} )^\dag 
+\bar{\alpha}_k ^* \hat{b}_{\bm k} ^{(\sigma)}  -\bar{\beta}_k (\hat{b}_{-\bm {k}}  ^{(\sigma)})^\dag,
\label{arewritten}
\\
\hat{b}_{\bm k} ^{(S)} &=\gamma^* _k \hat{a}_{\bm k} ^{(\phi)}  -\delta_k (\hat{a}_{- \bm {k}} ^{(\phi)} )^\dag
+\bar{\gamma}_k ^* \hat{b}_{\bm k} ^{(\sigma)} -\bar{\delta}_k (\hat{b}_{-\bm {k}}  ^{(\sigma)})^\dag,
\label{brewritten}
\end{align}
by solving the correspondence between the in- and out-creation/annihilation operators
\begin{equation}
\begin{pmatrix}\hat{a}_{\bm k} ^{(\phi)}  \\
(\hat{a}_{- \bm {k}} ^{(\phi)} )^\dag  \\
 \hat{b}_{\bm k} ^{(\sigma)} \\
(\hat{b}_{-\bm {k}}  ^{(\sigma)})^\dag\\
\end{pmatrix}=
\begin{pmatrix}\alpha_k & \beta_k & \gamma_k & \delta_k \\
\beta^* _k & \alpha^* _k & \delta^* _k & \gamma^* _k \\
\bar \alpha_k &\bar \beta_k  & \bar \gamma_k & \bar\delta_k \\
\bar \beta^* _k & \bar\alpha^*  _k& \bar \delta^* _k & \bar \gamma^* _k \\
\end{pmatrix}
\begin{pmatrix}\hat{a}_{\bm k} ^{(\Phi)} \\
(\hat{a}_{- \bm {k}} ^{(\Phi)} )^\dag \\
\hat{b}_{\bm k} ^{(S)} \\
(\hat{b}_{-\bm {k}}  ^{(S)})^\dag \\
\end{pmatrix},
\label{correspondence}
\end{equation}
in terms of $\hat{a}_{\bm k} ^{(\Phi)}$ and $\hat{b}_{\bm k} ^{(S)}$.
%Now we can define the out-vacuum state as the direct product of the Fock vacuum of $\hat{A}$ and $\hat{B}$ in the same way as the in-vacuum state,
%
%\begin{equation}
%\hat{A}_{\bm k}|0\rangle_A=0, \qquad
%\hat{B}_{\bm k}|0\rangle_{B}=0,  \qquad
%|0\rangle_{\rm out}\equiv |0\rangle_A \otimes |0\rangle_{B}\,.
%\label{out vac def}
%\end{equation}
%
%With eqs.~\eqref{arewritten} and \eqref{brewritten}, we can give the physical %meaning of the coefficients.
%For example, eq.~\eqref{arewritten} means that even in the state without %$\Phi$ particle,
%there exists $\phi$ particle unless $\beta_k =0$ and $\sigma$ particle unless %$\bar{\beta}_k =0$.
%In the limit $\Phi = \phi$, the former represents the particle production %as it gives
%Bogoliubov transformation, while the latter characterize the effect of entanglement %as it gives
%the mixing between the two fields. Therefore, we expect that $\beta_k$ and %$\bar{\delta}_k$
%represent the effect of particle production of $\phi$ and $\sigma$, respectively,
%while $\bar{\beta}_k$ and $\delta_k$ represent the effect of entanglement %between $\phi$ and $\sigma$. 
Plugging these expressions into eqs.~\eqref{commurelation early}, we find
\begin{align}
|\alpha_k|^2-|\beta_k|^2+|\bar\alpha_k|^2-|\bar\beta_k|^2&=D=1,
\\
|\gamma_k|^2-|\delta_k|^2+|\bar\gamma_k|^2-|\bar\delta_k|^2&=1,
\\
\alpha^* _k \delta_k -\beta_k \gamma^*_k+\bar\alpha^*_k \bar\delta_k -\bar\beta_k \bar\gamma^*_k&=0,
\label{cond3}
\\
\alpha^*_k \gamma_k-\beta_k\delta^*_k+\bar\alpha^*_k\bar\gamma_k-\bar\beta_k\bar\delta^*_k &=0,
\end{align}
where $D$ is the determinant of the $4\times 4$ matrix in eq.~(\ref{correspondence}).
%is given by
%
%\begin{equation}
%D\equiv \left|\begin{matrix}\alpha & \beta & \gamma & \delta \\
%\beta^* & \alpha^* & \delta^* & \gamma^* \\
%\bar \alpha &\bar \beta  & \bar \gamma & \bar\delta \\
%\bar \beta^* & \bar\alpha^* & \bar \delta^* & \bar \gamma^* \\
%\end{matrix}\right|=|\alpha|^2-|\beta|^2+|\bar\alpha|^2-|\bar\beta|^2.
%\end{equation}
%
One can show that these relations are equivalent to eqs.~\eqref{cond1}-\eqref{cond4}.

Next, making use of the relations between the two sets of creation/annihilation operators,
we will show how the in-vacuum state is related to 
the out-vacuum state.  
%For this purpose, we start with rewriting 
%the in-vacuum state in terms of the out-state as
Since the in-vacuum state should look like an excited state of the out-vacuum,
it can be written as
\begin{equation}
|0\rangle_{\rm in}= \prod_{\bm k} f_{\bm k}\left[(\hat{a}_{\bm k} ^{(\phi)} )^\dag,(\hat{a}_{- \bm k} ^{(\phi)} )^\dag,
(\hat{b}_{\bm k} ^{(\sigma)} )^\dag,(\hat{b}_{-\bm k} ^{(\sigma)} )^\dag\right]\,|0\rangle_{\rm out}
\equiv \hat{f} |0\rangle_{\rm out}\,,
\label{inoutvac}
\end{equation}
where $f_{\bm k} [(\hat{a}_{\bm k} ^{(\phi)} )^\dag,(\hat{a}_{- \bm k} ^{(\phi)} )^\dag,
(\hat{b}_{\bm k} ^{(\sigma)} )^\dag,(\hat{b}_{-\bm k} ^{(\sigma)} )^\dag]$ 
is a function of the creation operators.
Letting $\hat{a}_{\bm k} ^{(\Phi)} $ act on both sides of the above equation, 
from eq.~(\ref{arewritten}), one finds
\begin{align}
0&=
%\left[\alpha^* _k \hat{a}_{\bm {k}} ^{(\phi)} -\beta_k (\hat{a}_{-\bm {k}} ^{(\phi)} )^\dag
%+\bar{\alpha}^* _k \hat{b}_{\bm {k}} ^{(\sigma)} -\bar{\beta}_k (\hat{b}_{-\bm {k}} ^{(\sigma)} )^\dag \right]
%\prod_{\bm k'} f_{\bm k}((\hat{a}_{\bm {k}} ^{(\phi)} )^\dag,(\hat{a}_{-\bm {k}} ^{(\phi)} )^\dag,(\hat{b}_{\bm {k}} %^{(\sigma)} )^\dag,
%(\hat{b}_{-\bm {k}} ^{(\sigma)} )^\dag)\,|0\rangle_{\rm out}\,,
%\notag\\
%&=
\left[\alpha^* _k \frac{\partial \hat{f}}{\partial (\hat{a}_{\bm {k}} ^{(\phi)} )^\dag}
+\bar{\alpha}^* _k \frac{\partial \hat{f}}{\partial (\hat{b}_{\bm {k}} ^{(\sigma)} )^\dag }
-\left(\beta_k (\hat{a}_{-\bm {k}} ^{(\phi)} )^\dag+ \bar{\beta}_k (\hat{b}_{-\bm {k}} ^{(\sigma)} )^\dag\right)\hat{f}
\right]|0\rangle_{\rm out}\,,
\label{PDA}
\end{align}
where $[\hat{a}_{\bm k} ^{(\phi)} ,\hat{f}]=\partial \hat{f}/\partial (\hat{a}_{\bm k} ^{(\phi)} )^\dag$ 
and $[\hat{b}_{\bm k} ^{(\sigma)} ,\hat{f}]=\partial \hat{f}/\partial (\hat{b}_{\bm k} ^{(\sigma)} )^\dag$  are used.
Since eq.~\eqref{PDA} only depends on the creation operators, the parenthesis itself must vanish,
\begin{equation}
\alpha^*_k \frac{\partial \hat{f}}{\partial  (\hat{a}_{\bm {k}} ^{(\phi)} )^\dag}
+\bar{\alpha}^*_k\frac{\partial \hat{f}}{\partial  (\hat{b}_{\bm {k}} ^{(\sigma)} )^\dag}
-\left(\beta_k  (\hat{a}_{-\bm k} ^{(\phi)} )^\dag + \bar{\beta}_k  (\hat{b}_{-\bm k} ^{(\sigma} )^\dag\right) \hat{f}=0.
\label{f dif eq}
\end{equation}
This differential equation can be solved with an ansatz
\begin{eqnarray}
f_{\bm k}&=&N_{\bm k} ^{f} \exp\biggl[\frac12 \biggl( \mathcal{C}^{\phi \phi}_{k}\,
(\hat{a}_{-\bm k} ^{(\phi)} )^\dag (\hat{a}_{\bm k} ^{(\phi)} )^\dag
+\mathcal{C}^{\sigma \sigma}_{k}\,(\hat{b}_{-\bm k} ^{(\sigma)} )^\dag (\hat{b}_{\bm k} ^{(\sigma)} )^\dag
\nonumber\\
&&\qquad\qquad\qquad
+\mathcal{C}^{\phi \sigma}_{k}\left((\hat{a}_{-\bm k} ^{(\phi)} )^\dag (\hat{b}_{\bm k} ^{(\sigma)} )^\dag+
(\hat{a}_{\bm k} ^{(\phi)} )^\dag (\hat{b}_{-\bm k} ^{(\sigma)} )^\dag \right)\biggr)\biggr],
\label{expression fk}
\end{eqnarray}
where $N_{\bf k}^f$ is a normalization factor.
Noting that $ (\hat{a}_{\bm k} ^{(\phi)} )^\dag$, and $ (\hat{a}_{-\bm k} ^{(\phi)} )^\dag$ commute 
and plugging this ansatz into eq.~\eqref{f dif eq},
we obtain the relation
\begin{equation}
%\ln f=C_1  \hat{A}^\dag_{\bm k}+C_2 \hat{B}^\dag_{\bm k}+C_3,
%\qquad
\alpha^*_k \left( \mathcal{C}^{\phi \phi}_{k} (\hat{a}_{-\bm k} ^{(\phi)} )^\dag +  \mathcal{C}^{\phi \sigma}_{k} 
 (\hat{b}_{-\bm k} ^{(\sigma)} )^\dag\right)
 +\bar{\alpha}^*_k  \left( \mathcal{C}^{\sigma \sigma}_{k} (\hat{b}_{-\bm k} ^{(\sigma)} )^\dag 
 +  \mathcal{C}^{\phi \sigma}_{k} (\hat{a}_{-\bm k} ^{(\phi)} )^\dag\right)
 -\beta_k  (\hat{a}_{-\bm k} ^{(\phi)} )^\dag 
- \bar{\beta}_k (\hat{b}_{-\bm k} ^{(\sigma)} )^\dag =0,
\end{equation}
where we have used the fact that $\partial f_{\bm k'}/\partial (\hat{a}_{\bm k} ^{(\phi)} )^\dag$
and $\partial f_{\bm k'}/\partial (\hat{b}_{\bm k} ^{(\sigma)} )^\dag$ become nonzero
only for ${\bm k'} = {\bm k}$ and ${\bm k'} = -{\bm k}$.
Letting $\hat{b}_{\bm k}$ act on eq.~\eqref{inoutvac}, one also finds
\begin{equation}
\gamma^* _k  \left( \mathcal{C}^{\phi \phi}_{k} (\hat{a}_{-\bm k} ^{(\phi)} )^\dag +  \mathcal{C}^{\phi \sigma}_{k} 
 (\hat{b}_{-\bm k} ^{(\sigma)} )^\dag\right)
 +\bar{\gamma}^*_k \left( \mathcal{C}^{\sigma \sigma}_{k} (\hat{b}_{-\bm k} ^{(\sigma)} )^\dag 
 +  \mathcal{C}^{\phi \sigma}_{k} (\hat{a}_{-\bm k} ^{(\phi)} )^\dag\right)
-\delta_k  (\hat{a}_{-\bm k} ^{(\phi)} )^\dag  -
\bar{\delta}_k (\hat{b}_{-\bm k} ^{(\sigma)} )^\dag=0.
\end{equation}
We then obtain the coefficients $ \mathcal{C}^{\phi \phi}_{k}$, $\mathcal{C}^{\sigma \sigma}_{k}$ 
and $\mathcal{C}^{\phi \sigma}_{k}$ in eq.~\eqref{expression fk}  as
%
%\begin{equation}
%f=N_f\exp\left[\frac12 \left( \mathcal{C}_{\phi \phi k}\,
%(\hat{a}_{-\bm {k}} ^{(\phi)} )^\dag (\hat{a}_{\bm {k}} ^{(\phi)} )^\dag
%+\mathcal{C}^{\sigma \sigma}_{k}\,(\hat{b}_{-\bm {k}} ^{(\sigma)} )^\dag (\hat{b}_{\bm {k}} ^{(\sigma)} )^\dag
%+\mathcal{C}^{\phi \sigma}_{k}\left((\hat{a}_{-\bm {k}} ^{(\phi)} )^\dag (\hat{b}_{\bm {k}} ^{(\sigma)} )^\dag+
%(\hat{a}_{\bm {k}} ^{(\phi)} )^\dag (\hat{b}_{-\bm {k}} ^{(\sigma)} )^\dag \right)\right)\right],
%\end{equation}
%
%with
%
\begin{equation}
\mathcal{C}^{\phi \phi}_{k}=\frac{\beta_k\bar{\gamma}^*_k-\bar{\alpha}^*_k \delta_k}{\alpha^*_k \bar{\gamma}^*_k-\bar{\alpha}^*_k \gamma^*_k},
\qquad
\mathcal{C}^{\sigma \sigma}_{k}=\frac{\alpha^*_k\bar{\delta}_k-\bar{\beta}_k\gamma^*_k}{\alpha^*_k\bar{\gamma}^*_k-\bar{\alpha}^* _k \gamma^*_k},
\qquad
\mathcal{C}^{\phi \sigma}_{k}=\frac{\alpha^*_k\delta_k -\beta_k\gamma^*_k}{\alpha^*_k\bar{\gamma}_k^*-\bar{\alpha}^*_k \gamma^*_k},
\label{C parameters}
\end{equation}
where the expressions in eq.~\eqref{cond3} and $\bar{\alpha}^*_k \bar{\delta}_k-\bar{\beta}_k\bar{\gamma}^*_k=-\alpha^*_k\delta_k +\beta_k\gamma^*_k$, are used. Appendix~\ref{app:Bogcoeff} then perturbatively solves the relations (\ref{C parameters}) with respect to the generalized Bogoliubov coefficients ($\alpha_k$, $\beta_k$, $\gamma_k$, $\delta_k$, $\bar{\alpha}_k$, $\bar{\beta}_k$, $\bar{\gamma}_k$, $\bar{\delta}_k$) to show that there exists a family of generalized Bogoliubov coefficients that corresponds to ($C_k^{\phi\phi}$, $C_k^{\sigma\sigma}$, $C_k^{\phi\sigma}$).The relations \eqref{cond1}-\eqref{cond4} also imply the following identity.
\begin{equation}
\mathcal{C}^{\phi \phi}_k\mathcal{C}^{\sigma \sigma}_{k} - \mathcal{C}^{\phi \sigma}_k\mathcal{C}^{\phi \sigma}_k = \frac{\beta_k\bar\delta_k-\bar\beta_k\delta_k}{\alpha^*_k \bar\gamma^*_k - \bar\alpha^*_k \gamma^*_k}\,.
\end{equation}

The terms with $\mathcal{C}^{\phi \phi}_{k}$ and $\mathcal{C}^{\sigma\sigma}_{k}$
in the exponent in eq.~\eqref{expression fk} lead to a vacuum squeezed state
and can be understood as a generalization of the Bogoliubov transformation.
Indeed, they reproduce the conventional results, $\mathcal{C}^{\phi \phi}_{k}\to\beta_k/\alpha_k^*$
and $\mathcal{C}^{\sigma \sigma}_{k}\to \bar\delta_k/\bar\gamma_k^*$ , in the limit that the additional coefficients $\gamma_k, \delta_k, \bar\alpha_k$ and $\bar \beta_k$ vanish. 
Interestingly, however, if $C_k^{\phi\sigma}$ is non-vanishing,
the in-vacuum state acquires cross terms proportional to 
$(\hat{a}_{-\bm {k}} ^{(\phi)} )^\dag (\hat{b}_{\bm {k}} ^{(\sigma)} )^\dag+
(\hat{a}_{\bm {k}} ^{(\phi)} )^\dag (\hat{b}_{-\bm {k}} ^{(\sigma)} )^\dag $ in the exponent which lead to an entangled state. 

The entanglement of the state becomes more evident 
by checking the separability of the state. If $C_k^{\phi\sigma}=0$,
the in-vacuum state separates into the $\phi$ and $\sigma$ parts and hence the system is not entangled:
\begin{equation}
|0\rangle_{\rm in}\xrightarrow{C_k^{\phi\sigma}=0}
N_{\bm k}\left(\exp\left[\frac12  \mathcal{C}^{\phi \phi}_{k}\,
(\hat{a}_{-\bm k} ^{(\phi)} )^\dag (\hat{a}_{\bm k} ^{(\phi)} )^\dag\right]
 |0\rangle_\phi \right) 
\otimes 
\left(\exp\left[\frac12  \mathcal{C}^{\sigma \sigma}_{k}\,
(\hat{a}_{-\bm k} ^{(\sigma)} )^\dag (\hat{a}_{\bm k} ^{(\sigma)} )^\dag\right]
 |0\rangle_\sigma \right) 
\,.
\end{equation}
On the other hand, if $C_k^{\phi\sigma}\neq 0$, the in-vacuum state is no longer separable, and thus the system is entangled.
This is a unique feature of a system in which multiple fields are mixed.
We conclude that the quantum fluctuations of $\phi$ and $\sigma$ are entangled unless 
$\mathcal{C}^{\phi \sigma}_{k}$ 
%and $\bar{\alpha}^*\bar{\delta}-\bar{\beta}\bar{\gamma}^*$ 
vanishes,
\begin{equation} \label{entangle condition}
{\rm Entangle}  \quad\Longleftrightarrow\quad
\alpha^*_k \delta_k -\beta_k \gamma^*_k\neq 0.
%\quad{\rm or}\quad \bar{\alpha}^*\bar{\delta}-\bar{\beta}\bar{\gamma}^*\neq0.
\end{equation}
%

%Note that by  virtue of momentum conservation, 
%these operators in eq.~\eqref{expression fk} always cause  pair-productions %of particles whose total momenta vanish.

%====================================================================================%
\section{Entanglement in Schr\"odinger Picture\label{sec_schrodinger}} 
%====================================================================================%

In the previous section, we discussed the effect of particle production and entanglement
based on the Heisenberg picture.
%, where we considered the time evolution of $\hat{u}_{\phi} (\eta,\bm k)$ %and $\hat{u}_{\sigma} (\eta,\bm k)$ with fixed in-vacuum. 
On the other hand, in the previous work by Albrecht, Bolis and Holman
(ABH) \cite{Albrecht:2014aga}, the effect of entanglement was analyzed within the 
Schr\"odinger picture\footnote{For other on the use of the Schr\"odinger picture see \cite{Boyanovsky:1993xf,  Anderson:2005hi, Freese:1984dv}.} . In this work,  the time evolution of  the system is encoded in a wave functional of the fields.
To bridge between these two approaches, in this section, we translate the results of the previous section
into the language of the Schr\"odinger picture
and compare it with the ABH state. 
We use $\delta \phi$ and $\delta \sigma$ instead of $u_\phi$ and $u_\sigma$ and, since it is obvious that we are considering the perturbations of $\phi$ and $\sigma$, we adopt the
 shorthand notation $\delta \phi \to \phi$
 and $\delta \sigma \to \sigma$ in this section.

%====================================================================================%
\subsection{Out-vacuum wave function}
%====================================================================================%

Here, we obtain the expression of the wave function of the out-vacuum.
Based on the late-time decoupled Lagrangian eq.~\eqref{late time action},
the conjugate momenta of the fields in Fourier space are given by
\begin{equation}
\Pi^\phi(\eta,\bm{k})\equiv \frac{\partial S}{\partial \phi'_{\bm k}}=a^2\phi'(\eta,-\bm{k}),
\qquad
\Pi^\sigma(\eta,\bm{k})\equiv \frac{\partial S}{\partial \sigma'_{\bm k}}=a^2\sigma'(\eta,-\bm{k}).
\end{equation}
At late times, where both $G_{IJ}$ and $M_{IJ}$ are diagonalized as eq.~\eqref{GM diagonalized late}, 
the creation/annihilation operators associated with the out-vacuum can be rewritten in terms of these original fields and their conjugate momenta as
\begin{align}
 \begin{pmatrix}\hat{a}_{-\bm {k}} ^{(\phi)} \\ (\hat{a}_{-\bm {k}} ^{(\phi)} )^\dag \\ \end{pmatrix}
=-i \begin{pmatrix}a^2(u_{\phi k}^*/a)' & -u_{\phi k}^*/a \\ -a^2(u_{\phi k}/a)' & u_{\phi k}/a \\\end{pmatrix}
\begin{pmatrix}\hat{\phi}_{\bm k} \\ \hat{\Pi}^\phi_{-\bm{k}} \\ \end{pmatrix},
\label{A fields}
\end{align}
\begin{align}
\begin{pmatrix}\hat{b}_{\bm {k}} ^{(\sigma)}   \\  (\hat{b}_{-\bm {k}} ^{(\sigma)} )^\dag \\ \end{pmatrix}
=-i \begin{pmatrix}a^2(u_{\sigma k}^*/a)' & -u_{\sigma k}^*/a \\ -a^2(u_{\sigma k}/a)' & u_{\sigma k}/a \\\end{pmatrix}
\begin{pmatrix}\hat{\sigma}_{\bm k} \\ \hat{\Pi}^\sigma_{-\bm{k}} \\ \end{pmatrix},
\label{B fields}
\end{align}
where $u_{\phi k}$ and $u_{\sigma k}$ are 
%the ones for Bunch-Davies vacuum
given by 
eqs.~(\ref{uphi_BD}) and (\ref{usigma_BD}),
%. since we are interested in the out-vacuum 
and  we used the Wronskian relations (\ref{mode equation late}).
%%
%\begin{equation}
%u_\varphi {u_\varphi^*}' -u_\varphi' u_\varphi^*=i, \qquad
%u_s {u_s^*}' -u_s' u_s^*=i.
%\end{equation}
%%
The equal-time commutation relations of these fields, 
$[\phi(\eta,\bm{x}),\Pi^\phi(\eta,\bm{y})]=i\delta(\bm{x}-\bm{y})$ and
$[\sigma(\eta,\bm{x}),\Pi^\sigma(\eta,\bm{y})]=i\delta(\bm{x}-\bm{y})$,
are recast in Fourier space as
\begin{equation}
[\phi_{\bm k}, \Pi^\phi_{\bm p}]=i(2\pi)^3 \delta(\bm k+\bm p),
\qquad
[\sigma_{\bm k}, \Pi^\sigma_{\bm p}]=i(2\pi)^3 \delta(\bm k+\bm p).
\end{equation}
Thus, the conjugate momenta of an eigenstate of $\hat{\phi}_{\bm{k}}(\eta)$ and $\hat{\sigma}_{\bm{k}}(\eta)$, for each wavenumber $\bm k$, $|\{\phi_{\bm k}, \sigma_{\bm k}\}(\eta)\rangle$, are defined in terms of the derivatives with respect to the fields,
\begin{align}
\hat{\Pi}^\phi_{\bm k}|\{\phi_{\bm k}, \sigma_{\bm k}\}\rangle
&=-i(2\pi)^3\frac{\partial}{\partial \phi_{\bm k}}|\{\phi_{\bm k}, \sigma_{\bm k}\}\rangle,
\qquad
\hat{\Pi}^\sigma_{\bm k}|\{\phi_{\bm k}, \sigma_{\bm k}\}\rangle
=-i(2\pi)^3\frac{\partial}{\partial \sigma_{\bm k}}|\{\phi_{\bm k}, \sigma_{\bm k}\}\rangle.
\end{align}
Letting $\langle \{\phi_{\bm k}, \sigma_{\bm k}\}|$  act on eq.~\eqref{out vac def}, one finds\footnote{One can show that $\langle \{\phi_{\bm k}, \sigma_{\bm k}\}|\hat{\phi}_{\pm \bm{k}}=
\phi_{\pm \bm{k}}\langle \{\phi_{\bm k}, \sigma_{\bm k}\}|$ and 
$\langle \{\phi_{\bm k}, \sigma_{\bm k}\}| \hat{\Pi}^\phi_{\pm \bm k}=-i(2\pi)^{-3}(\partial/\partial \phi_{\pm\bm k})\langle \{\phi_{\bm k}, \sigma_{\bm k}\}|$ with analogous equations for $\sigma_{\pm \bm k}$ by replacing $\phi_{\pm \bm k}\to \sigma_{\pm \bm k}$.}
\begin{equation}
\langle \{\phi_{\bm k}, \sigma_{\bm k}\}| \hat{a}_{\bm {k}} ^{(\phi)} |0\rangle_\phi=
\left(-i a^2\left(\frac{u_{\phi k}^*}{a}\right)' \phi_{\bm k}+(2\pi)^3 \frac{u_{\phi k}^*}{a}
\frac{\partial}{\partial \phi_{-\bm k}}\right)
\langle \{\phi_{\bm k}, \sigma_{\bm k}\}|0\rangle_\phi=0,
\end{equation}
with an analogous equation for $\sigma$ in which one replaces $\phi \leftrightarrow \sigma$.
Substituting the following Gaussian wave function for 
the out-vacuum, with a normalization factor $\mathcal{N} ^{({\rm out})}$
\begin{equation}
\langle \{\phi_{\bm k}, \sigma_{\bm k}\}(\eta)|0\rangle_{\rm out}
=\mathcal{N} ^{({\rm out})}(\eta)  \exp\left[-\frac{1}{2}\int\frac{\dd^3 k}{(2\pi)^3}\Big(\omega^\phi_k(\eta) \phi_{-\bm k} \phi_{\bm k}+\omega^\sigma_k (\eta) \sigma_{-\bm k} \sigma_{\bm k}\Big) \right],
\end{equation} 
we find 
\begin{equation}
\omega^\phi_k(\eta)= -i a^{2}\partial_\eta \ln(u_{\phi k}^*/a),
\qquad
\omega^\sigma_k(\eta)= -ia^{2} \partial_\eta \ln(u_{\sigma  k}^*/a).
\label{omega sols}
\end{equation}
{Note that this wave function is valid only at late times and
one needs to use the full-Hamiltonian to find the wave function at earlier times.}
For each wave number $\bm k$ the wave function is
\begin{equation}
\langle \phi_{\bm k}, \sigma_{\bm k}|0\rangle_{\rm out}
=\mathcal{N}_{\bm k} ^{({\rm out})} \exp\left[ -\frac{1}{2}\left(\omega^\phi_k\, \phi_{-\bm k} \phi_{\bm k}+\omega^\sigma_k\, \sigma_{-\bm k} \sigma_{\bm k}\right) \right],
\end{equation} 
where $\mathcal{N}_{\bm k} ^{({\rm out})} $ is a normalization factor. One can see that the $\phi$ and $\sigma$ sectors of the wave function for the out-vacuum are separable, and therefore this state is not entangled. This is to be expected, as $\phi$ and $\sigma$ are in the BD-vacuum at late times, however, as we shall see in the following section, this is not the case for the in-vacuum state. 

%====================================================================================%
\subsection{In-vacuum wave function}
%====================================================================================%

We can  obtain the wave function of the in-vacuum state in a similar way to the out-vacuum case. Substituting eqs.~\eqref{A fields} and \eqref{B fields}  into eqs.~\eqref{arewritten}
and \eqref{brewritten},
one finds
\begin{align}
\hat{a}_{\bm k} ^{\Phi}&=-ia^2 F_{\phi k}'(\alpha_k,\beta_k)\hat{\phi}_{\bm k}
+iF_{\phi k}(\alpha_k,\beta_k)\hat{\Pi}_{-\bm k}^\phi
-ia^2 F_{\sigma k}'(\bar\alpha_k,\bar\beta_k)\hat{\sigma}_{\bm k}
+iF_{\sigma k}(\bar\alpha_k,\bar\beta_k)\hat{\Pi}_{-\bm k}^\sigma,
\\
\hat{b}_{\bm k} ^{S}&= -ia^2 F_{\phi k}'(\gamma_k,\delta_k)\hat{\phi}_{\bm k}
+iF_{\phi k}(\gamma_k,\delta_k)\hat{\Pi}_{-\bm k}^\phi
-ia^2 F_{\sigma k}'(\bar\gamma_k,\bar\delta_k)\hat{\sigma}_{\bm k}
+iF_{\sigma k}(\bar\gamma_k,\bar\delta_k)\hat{\Pi}_{-\bm k}^\sigma,
\end{align}
where we have defined
\begin{equation}
F_{I k}(x,y)\equiv a^{-1}(\eta)\big[x^* u_{z k}^*(\eta)+y\, u_{z k}(\eta)\big],
\qquad (I=\phi, \sigma),
\end{equation}
and $F'_{I k}(x,y)\equiv \partial_\eta F_{I k}(x,y)$.
Plugging a Gaussian expression with a cross term and normalization factor $\mathcal{N}^{({\rm in})}$
\begin{multline}
\langle \{\phi_{\bm k}, \sigma_{\bm k}\}{(\eta)}|0\rangle_{\rm in}
\\=\mathcal{N}^{({\rm in})}{(\eta)} \exp\left[-\frac{1}{2}\int\frac{\dd^3 k}{(2\pi)^3}\Big\{\Omega^\phi_k{(\eta)} \phi_{-\bm k} \phi_{\bm k}+\Omega^\sigma_k{(\eta)} \sigma_{-\bm k} \sigma_{\bm k}+\Omega_k^{\phi\sigma}{(\eta)}\left(\phi_{-\bm k}\sigma_{\bm k}+\sigma_{-\bm k}\phi_{\bm k}\right)\Big\} \right],
\end{multline}
into 
\begin{equation}
\langle \{\phi_{\bm k}, \sigma_{\bm k}\}|\hat{a}_{\bm k}|0\rangle_{\rm in} =0,
\qquad
\langle \{\phi_{\bm k}, \sigma_{\bm k}\}|\hat{b}_{\bm k}|0\rangle_{\rm in} =0,
\end{equation}
we obtain the following four equations
\begin{equation}
\begin{pmatrix}
F_{\phi k}(\alpha_k,\beta_k) & 0 & F_{\sigma k}(\bar\alpha_k,\bar\beta_k) \\
0 & F_{\sigma k}(\bar\alpha_k,\bar\beta_k) & F_{\phi k}(\alpha_k,\beta_k) \\
F_{\phi k}(\gamma_k,\delta_k) & 0 & F_{\sigma k}(\bar\gamma_k,\bar\delta_k) \\
0 & F_{\sigma k}(\bar\gamma_k,\bar\delta_k) & F_{\phi k}(\gamma_k,\delta_k) \\
\end{pmatrix}
\begin{pmatrix}
\Omega_k^\phi \\
\Omega_k^\sigma \\
\Omega_k^{\phi\sigma} \\
\end{pmatrix}=
-ia^2\begin{pmatrix}
F_{\phi k}'(\alpha_k,\beta_k) \\
F_{\sigma k}'(\bar\alpha_k,\bar\beta_k) \\
F_{\phi k}'(\gamma_k,\delta_k) \\
F_{\sigma k}'(\bar\gamma_k,\bar\delta_k) \\
\end{pmatrix},
\end{equation}
where one equation out of four is redundant.
Solving these equations, we find
%
%\begin{align}
%\Omega^\phi_k =i a^{2}\partial_\eta^{\varphi} \ln\left[U/a\right],
%\qquad
%\Omega^\sigma_k =i a^{2}\partial_\eta^{s} \ln\left[U/a\right],
%\qquad
%\Omega^{\phi\sigma}_k= ia^2(\alpha^* \delta-\beta \gamma^*)/U,
%\label{Omega Sol}
%\end{align}
%
%

\begin{align}
\Omega^\phi_k(\eta) &=-i a^{2}\left[\partial_{\eta_1} \big[\ln U_k(\eta_1,\eta_2)\big]\Big|_{\eta_1=\eta_2=\eta} -\frac{a'}{a} \right],
\label{Omega Sol1}
\\
\Omega^\sigma_k(\eta) &=-i a^{2}\left[\partial_{\eta_2} \big[\ln U_k(\eta_1,\eta_2)\big]\Big|_{\eta_1=\eta_2=\eta} -\frac{a'}{a} \right],
\\
\Omega^{\phi\sigma}_k(\eta) &=- a^2 \mathcal{C}^{\phi \sigma}_k/U_k(\eta,\eta),
\label{Omega Sol}
\end{align}
with
\begin{align}
 U_k (\eta_1, \eta_2) \equiv 
&
u_{\phi k}^*(x_1)  u_{\sigma k}^*(x_2) 
+\mathcal{C}^{\phi \phi}_k u_{\phi k}(x_1)  u_{\sigma k}^*(x_2)\notag\\
& 
 +\mathcal{C}^{\sigma \sigma}_{k} u_{\phi k}^*(x_1)  u_{\sigma k}(x_2) +
\left( \mathcal{C}^{\phi \phi}_k\mathcal{C}^{\sigma \sigma}_{k} - \mathcal{C}^{\phi \sigma}_k\mathcal{C}^{\phi \sigma}_k\right)
u_{\phi k}(x_1) u_{\sigma k}(x_2)\,,
\label{definition U}
\end{align}
where $x_1\equiv -k\eta_1$ and $x_2\equiv -k\eta_2$.
In eq.~\eqref{definition U}, time dependence on $\eta_1$ and $\eta_2$ 
is introduced so that $u_{\phi k}$ depends on $\eta_1$, 
while $u_{\sigma k}$ depends on $\eta_2$. In eqs.~(\ref{Omega Sol}) after taking the derivatives with respect to 
$\eta_1$ and $\eta_2$, we set $\eta_1=\eta_2 = \eta$.
%$\partial_\eta^{\varphi}$ and $\partial_\eta^{s}$ in eq.~\eqref{Omega Sol} does not hit $u_s$ and $u_\varphi$, respectively (i.e. $\partial_\eta^{\varphi}$
%hits  $u_\varphi$ and the scale factor but not $u_s$).
In the limit $\alpha_k\to1, \bar\gamma_k\to1$, where all the other coefficients vanish and the in-vacuum and out-vacuum coincide, the above result reduces to eq.~\eqref{omega sols} as expected.

On the other hand, a non-zero $\Omega_k^{\phi\sigma}$ corresponds to an entangled state since the $\phi$ and $\sigma$ sectors would no longer be separable (see eq.~\eqref{entangle condition}).  The in-vacuum state of  $\phi$ and $\sigma$ is therefore entangled thanks to the evolution of the kinetic and mass matrix terms.
Focusing on a single wavenumber, 
the wave function of the in-vacuum state is given by
\begin{equation}
\langle \phi_{\bm k}, \sigma_{\bm k}|0\rangle_{\rm in}
=\mathcal{N}_{\bm k} ^{({\rm in})} \exp\left[-\frac{1}{2}\Omega^\phi_k\, \phi_{-\bm k} \phi_{\bm k}-\frac{1}{2}\Omega^\sigma_k\, \sigma_{-\bm k} \sigma_{\bm k} 
-\frac{1}{2}\Omega_k^{\phi\sigma}\left(\phi_{-\bm k}\sigma_{\bm k}+\sigma_{-\bm k}\phi_{\bm k}\right)\right],
\label{exact wave function}
\end{equation} 
where $\mathcal{N}_{\bm k} ^{({\rm in})}$ is a normalization factor.  
{It can be shown that this wave function satisfies the Schr\"odinger equation with the free Hamiltonian constructed from the late-time
Lagrangian eq.~\eqref{late time action} as expected. However, it should be noted that the above wave function with eqs.~\eqref{Omega Sol1}-\eqref{definition U} is valid only at late times ($t\geq t_0$) when the Hamiltonians for  $\phi$ and $\sigma$ are decoupled, and one needs to solve the Schr\"odinger equation with the full-Hamiltonian to obtain the wave function at an earlier time (see Appendix \ref{app:Schrodinger} for further discussion based on 
a similar model as that discussed in Sec.\ref{sec_concrete model}).}

%====================================================================================%
\subsection{Comparison to {the} entangled state in ABH } 
%====================================================================================%

In this subsection we compare our result eq.~\eqref{exact wave function} with the entangled state used in the ABH paper (see eq.~(2.6) in \cite{Albrecht:2014aga}). In this work a Gaussian entangled state ansatz was used to phenomenologically test for small deviations from a Bunch Davies initial state. By working in the Schr\"{o}dinger quantum field theory picture, the dynamics of the parameters $A_k, B_k$ and $C_k$ of this wave-functional,
\begin{equation}
\psi_k[\phi_k, \chi_k; \eta]
=N_k(\eta)\exp\left[-\frac{1}{2}\left(A_k(\eta)\phi_{\vec k}\phi_{-\vec k}
+B_k(\eta)\chi_{\vec k}\chi_{-\vec k}+C_k(\eta)\left(\phi_{\vec k}\chi_{-\vec k}+\chi_{\vec k}\phi_{-\vec k}\right)\right)\right],
\label{abhstate}
\end{equation}
are governed by the Schr\"{o}dinger equation {with the free Hamiltonian
corresponding to eq.~\eqref{late time action}}. 
%and can be used to solve for the evolution of the mode functions of the %inflaton fluctuations entangled to a spectator field. 
Several interesting and distinguishing observational features arise from such a state, including small oscillations in the inflaton power spectrum  \cite{Albrecht:2014aga}.
The correspondence between eq.~\eqref{abhstate} and our result for the entangled in-vacuum state eq.~\eqref{exact wave function} is as follows:
\begin{equation}
A_k \leftrightarrow  \Omega ^{\phi}_k , \quad B_k \leftrightarrow \Omega ^{\sigma}_k , \quad C_k \leftrightarrow  \Omega ^{\phi \sigma}_k 
\end{equation}
Note that the parameters $A_k, B_k$ and $C_k$ are not simple functions of the Bunch-Davies mode functions, they are solutions to the Schr\"{o}dinger equation, and encode the dynamics of the mixing of the two fields.  In terms of the entangled mode functions $f_k, g_k$ in \cite{Albrecht:2014aga} they are:
\begin{equation}
A_k= -i a^2\left(\frac{f_k^{\prime}}{f_k} - \frac{a^{\prime}}{a}\right), \quad B_k =-i a^2\left(\frac{g_k^{\prime}}{g_k} - \frac{a^{\prime}}{a}\right), \quad C_k = a^2 \frac{\lambda}{f_k g_k}\,.
\end{equation}
In \cite{Albrecht:2014aga}, in order to directly compare to the standard Bunch-Davies results, they adopt the following initial conditions at $\eta=\eta_0$:
\begin{align}
&A_k(\eta_0)=\left. -i a^2 \partial_\eta\ln\left(\frac{f_k^{BD}}{a}\right)\right|_{\eta=\eta_0},
\quad
B_k(\eta_0)=\left.-i a^2 \partial_\eta\ln\left(\frac{g_k^{BD}}{a}\right)\right|_{\eta=\eta_0},
& ({\rm eq.(2.9)\ in \ ABH})
\\
&C_k(\eta_0)=\left.a^2 \frac{\lambda_k}{f_k^{BD}g_k^{BD}}\right|_{\eta=\eta_0},
& ({\rm eq.(2.11)\ in \  ABH})
\\
&f_k^{BD}=u_{\phi k}^*,\qquad g_k^{BD}=u_{\sigma k}^*,
& ({\rm eq.(2.17)\&(3.15)\ in \  ABH})
\end{align}
and hence, in the notation used in this work,
\begin{align}
&A_k(\eta_0)=\left.-i a^2 \partial_\eta\ln(u_{\phi k}^*/a)\right|_{\eta=\eta_0},
\qquad B_k(\eta_0)=\left.-i a^2 \partial_\eta\ln(u_{\sigma k}^*/a)\right|_{\eta=\eta_0}, \nonumber\\
& C_k(\eta_0)=\left.a^2\lambda_k(u_{\phi k}^* u_{\sigma k}^*)^{-1}\right|_{\eta=\eta_0},
\label{init_condit}
\end{align} 
where $\lambda_k$ is a real constant and sets the strength of the entanglement. Our in-vacuum state at $\eta=\eta_0$ coincides with the ABH state
%the initial state of ABH { at $\eta=\eta_0$}, 
if the following conditions are satisfied: 
\begin{align}
 & \mathcal{C}^{\phi \phi}_{k}=
 \left.
 \frac{\lambda_k^2u_{\phi k}^*u_{\sigma k}}
 {u_{\phi k}^*u_{\sigma k}^*-\lambda_k^2u_{\phi k}u_{\sigma k}}\right|_{\eta=\eta_0}\,,
 \qquad
 \mathcal{C}^{\sigma \sigma}_{k}=
 \left.
 \frac{\lambda_k^2u_{\phi k}u_{\sigma k}^*}
 {u_{\phi k}^*u_{\sigma k}^*-\lambda_k^2u_{\phi k}u_{\sigma k}}\right|_{\eta=\eta_0}\,,
 \nonumber\\
 & 
 \mathcal{C}^{\phi \sigma}_{k}=
 \left.
 \frac{-\lambda_ku_{\phi k}^*u_{\sigma k}^*}
 {u_{\phi k}^*u_{\sigma k}^*-\lambda_k^2u_{\phi k}u_{\sigma k}}\right|_{\eta=\eta_0}\,.
 \label{Cabh}
 \end{align}
One then finds that 
their entanglement parameter $\lambda_k$ is closely related to our $\mathcal{C}_k^{\phi\sigma}$,
\begin{equation}
\lambda_k \ne 0 \leftrightarrow \mathcal{C}_{k}^{\phi\sigma} \ne 0\,.
\end{equation}
Therefore the condition for the non-zero entanglement $C_k\neq 0$, used in  \cite{Albrecht:2014aga} ,
is equivalent to the condition $\mathcal{C}^{\phi \sigma}_{k}\neq 0$ in this work.

As shown in Appendix~\ref{app:Bogcoeff}, one can perturbatively find a family of generalized Bogoliubov coefficients ($\alpha_k$, $\beta_k$, $\gamma_k$, $\delta_k$, $\bar{\alpha}_k$, $\bar{\beta}_k$, $\bar{\gamma}_k$, $\bar{\delta}_k$) that correspond to a given values of ($C_k^{\phi\phi}$, $C_k^{\sigma\sigma}$, $C_k^{\phi\sigma}$). Combining this general result with (\ref{Cabh}), we conclude that for a given ABH state parameterized by $\lambda_k$, there exists a family of corresponding generalized Bogoliubov coefficients.

%====================================================================================%
\section{Concrete Example with Entangled State from Kinetic Mixing 
\label{sec_concrete model}}
%====================================================================================%
%\label{Concrete Example with Entangled In-vacuum State from Kinetic Mixing}

In this section, we consider a simple example of the scenario discussed in the previous sections with a sudden change in the  kinetic matrix of the scalar fields.
We show analytically that an entangled state is generated and 
confirm that  oscillations are produced 
in the power spectrum of  the inflaton perturbations.

%====================================================================================%
\subsection{Model description}
%====================================================================================%
\label{Model}
Here, we consider a toy model with no  mass-mixing and in which only $G_{IJ}$ depends on time through $f(t)$, 
%and mass-mixing is absent,
%
\begin{equation}
G_{IJ}(t)=\begin{pmatrix}1 & f(t) \\
f(t) & 1 \\
\end{pmatrix},
\qquad
M_{IJ}=\begin{pmatrix}0 & 0 \\
0 & m_\sigma^2 \\
\end{pmatrix}.
\label{example GIJ}
\end{equation}
For simplicity, we assume that $f(t)$ is a constant $f_c$ which suddenly vanishes at a certain time $\eta_*$,
\begin{equation}
f(\eta)=f_c\, \Theta(\eta_*-\eta),
\qquad
(0<f_c<1)
\label{example ft}
\end{equation}
where $\Theta(\eta)$ is the Heaviside function. 
The matrix $\mcK_{IJ}$ introduced in eq.~(\ref{relation two sets of asymptotic fields}) is given by 
\begin{equation}
\begin{pmatrix}\phi \\
\sigma \\
\end{pmatrix}= \mcK \begin{pmatrix} \Phi \\ S \\\end{pmatrix},
\qquad
\mcK\equiv 
\begin{pmatrix}1 & \frac{-f_c}{\sqrt{1-f_c^2}} \\
0 & \frac{1}{\sqrt{1-f_c^2}} \\
\end{pmatrix},
\label{K example}
\end{equation}
and the mass matrix in eq.~\eqref{past action} is given by
\begin{equation}
\tilde{M}_{IJ}=\begin{pmatrix}
0 & 0 \\
0 & \frac{m_\sigma^2}{1-f_c^2} \\
\end{pmatrix}.
\end{equation}
Thus the solutions of the mode functions which connect to the Bunch-Davies vacuum in the sub-horizon limit are
\begin{align}
u_{\Phi k}(\eta)&=\sqrt{\frac{\pi x}{4k}}H_{3/2}^{(1)}(x),
\label{Phik}
\\
u_{S k}(\eta)&=\sqrt{\frac{\pi x}{4k}}e^{-\frac{i}{2}\pi\nu_c}H_{\nu_c}^{(1)}(x),
\qquad \nu_c\equiv \sqrt{\frac{9}{4}-\frac{m_\sigma^2/H^2}{1-f_c^2}}.
\label{Sk}
\end{align}
%
%====================================================================================%
\subsection{Entanglement from multi-field dynamics}
%====================================================================================%
To obtain the coefficients $\alpha_k, \beta_k, \gamma_k, \delta_k, \bar\alpha_k, \bar\beta_k, \bar\gamma_k,$ and $\bar\delta_k$ introduced in eqs.~\eqref{A expression} and \eqref{B expression}, 
one needs to connect $u_{\Phi k}$ and $u_{S k}$ in eqs.~(\ref{Phik}) and (\ref{Sk})  
to $u_{\phi k}$ and $u_{\sigma k}$ in eqs.\eqref{uphi_BD} and \eqref{usigma_BD}.
 Therefore, we consider the matching condition
for these two sets of mode functions here.
Taking into account the time evolution of $f(\eta)$, one finds the action as
\begin{align}
S=\frac{1}{2}\int \dd \eta\, \dd^3& x \bigg[ 
u_{\phi}'^2-(\partial_i u_\phi)^2+\frac{a''}{a} u_\phi^2
+u_\sigma'^2-(\partial_i u_\sigma)^2+\left(\frac{a''}{a}-a^2m_\sigma^2 \right) u_\sigma^2
\notag\\
&\quad+2f\left\{ u_\phi' u_\sigma'-\partial_i u_\phi \partial_i u_\sigma +\left(\frac{a''}{a}+\frac{a'f'}{af}\right)
u_\phi u_\sigma
\right\}
 \bigg],
\end{align}
where the second line vanishes for $\eta>\eta_*$.
The equations of motion for the mode functions, $u_{\phi k}(\eta)$ and $u_{\sigma k}$ (k), are written as
\begin{equation}\partial_\eta\left[
\begin{pmatrix}1 & f(\eta) \\
f(\eta) & 1 \\
\end{pmatrix}
\begin{pmatrix} u_{\phi k}' \\
u_{\sigma k}' \\
\end{pmatrix}\right]
=
-\frac{f'(\eta)}{\eta}\begin{pmatrix}0 & 1 \\
1 & 0 \\
\end{pmatrix}
\begin{pmatrix}u_{\phi k} \\
u_{\sigma k} \\
\end{pmatrix}+\cdots,
\end{equation}
where $\cdots$ does not include the time derivative of $f(\eta)$.
Since $f'(\eta)=-f_c\,\delta(\eta-\eta_*)$,
the junction conditions are,
\begin{equation}
\begin{pmatrix}u_{\phi k} \\
u_{\sigma k} \\
\end{pmatrix}_+
=\mcK \begin{pmatrix}u_{\Phi k} \\
u_{S k} \\
\end{pmatrix}_-,
\qquad
\partial_x\begin{pmatrix} u_{\phi k} \\
u_{\sigma k} \\
\end{pmatrix}_+
=\left[\begin{pmatrix}1 & f_c \\
f_c & 1 \\
\end{pmatrix}\mcK \partial_x
+\frac{f_c}{x_*}\begin{pmatrix}0 & 1 \\
1 & 0 \\
\end{pmatrix}\mcK\right] \begin{pmatrix}u_{\Phi k} \\
u_{S k}\\
\end{pmatrix}_-\,,
\end{equation}
where the subscripts $+$ and $-$ denote the time $\eta=\eta_*\pm\Delta\eta$ with the limit $\Delta \eta\to 0$, and $x_*\equiv -k\eta_*$.
Substituting $\hat{u}_\Phi(\eta,\bm k), \hat{u}_S(\eta,\bm k), \hat{u}_\phi(\eta,\bm k)$ and 
$\hat{u}_\sigma(\eta,\bm k)$ 
given by eqs.~\eqref{uPhi expansion}, \eqref{uS expansion}, \eqref{uphi expansion} and \eqref{usigma expansion}
 into the above junction conditions, we can calculate the coefficient of each operator.
 For the coefficients of $\hat{a}_{\bm k} ^{(\phi)}$ in eq.~\eqref{A expression}, we obtain 
\begin{align}
\alpha_k&=1,\qquad \beta_k
=0,\\
\gamma_k&=\frac{f_c\, e^{-\frac{i}{2}\pi\nu_c}}{x_*\sqrt{1-f_c^2}}
\frac{\left[x_* H^{(2)}_{1/2} +3 H^{(2)}_{3/2} -x_* H^{(2)}_{5/2} \right]H^{(1)}_{\nu_c} }
{\left[H^{(1)}_{1/2} -H^{(1)}_{5/2} \right]H^{(2)}_{3/2} 
+\left[H^{(2)}_{5/2} -H^{(2)}_{1/2} \right]H^{(1)}_{3/2} },
\\
\delta_k^*&=-\frac{f_c\, e^{-\frac{i}{2}\pi\nu_c}}{x_*\sqrt{1-f_c^2}}
\frac{\left[x_* H^{(1)}_{1/2} +3 H^{(1)}_{3/2} -x_* H^{(1)}_{5/2} \right]H^{(1)}_{\nu_c} }
{\left[H^{(1)}_{1/2} -H^{(1)}_{5/2} \right]H^{(2)}_{3/2} 
+\left[H^{(2)}_{5/2} -H^{(2)}_{1/2} \right]H^{(1)}_{3/2} },
\end{align}
where the second argument of all the Hankel functions is $x_*$.
At this stage, we can see that since these coefficients satisfy the condition \eqref{entangle condition},
the in-vacuum state is entangled. 
We find that the expressions of $\bar\alpha_k, \bar\beta_k, \bar\gamma_k,$ and $\bar\delta_k$, which are the coefficients of $\hat{b}_{\bm k} ^{(\sigma)}$ in eq.~\eqref{B expression},
depend on $m_\sigma$, because the complex conjugate of 
$H_{\nu_\sigma}^{(1)}(x)$ is $H_{\nu_\sigma}^{(2)}(x)$ for $m_\sigma^2<9H^2/4$,
while it is $H_{-\nu_\sigma}^{(2)}(x)$ for  $m_\sigma^2>9H^2/4$.
If $m_\sigma^2<9H^2/4$ (i.e. $\nu_\sigma>0$). One finds
\begin{align}
\bar\alpha_k&=-f_c\, e^{i x_*}\sqrt{\frac{\pi x_*}{8}} H^{(2)}_{\nu_\sigma} ,
\qquad
\bar\beta^* _k=f_c\, e^{i x_*}\sqrt{\frac{\pi x_*}{8}} H^{(1)}_{\nu_\sigma} ,
\\
\bar \gamma_k&=\frac{i\pi}{8\sqrt{1-f_c^2}}\bigg[
2(1-f_c^2)\left(H_{3/2}^{(1)} -x_* H_{1/2}^{(1)} \right)H_{\nu_\sigma}^{(2)} \\
&\qquad\qquad\qquad+e^{-\frac{i}{2}\pi \nu_c}\left((1+2f_c^2+2\nu_\sigma)H_{\nu_\sigma}^{(2)}-2x_* H_{1+\nu_\sigma}^{(2)}\right) H^{(1)}_{\nu_c}
\bigg],
\\
\bar \delta^*_k&=\frac{-i\pi}{8\sqrt{1-f_c^2}}\bigg[
2(1-f_c^2)\left(H_{3/2}^{(1)} -x_* H_{1/2}^{(1)} \right)H_{\nu_\sigma}^{(1)} \\
&\qquad\qquad\qquad+e^{-\frac{i}{2}\pi \nu_c}\left((1+2f_c^2+2\nu_\sigma)H_{\nu_\sigma}^{(1)}-2x_* H_{1+\nu_\sigma}^{(1)}\right) H^{(1)}_{\nu_c}
\bigg],
\end{align}
while if $m_\sigma^2>9H^2/4$ (i.e. $\nu_\sigma$ is imaginary), one obtains
\begin{align}
\bar\alpha_k&=-f_c\, e^{i (x_*+\pi\nu_\sigma)}\sqrt{\frac{\pi x_*}{8}} H^{(2)}_{-\nu_\sigma} ,
\qquad
\bar\beta^*_k=f_c\, e^{i(x_*+\pi\nu_\sigma)}\sqrt{\frac{\pi x_*}{8}} H^{(1)}_{\nu_\sigma} ,
\\
\bar \gamma_k&=\frac{i\pi\, e^{i \pi \nu_\sigma}}{8\sqrt{1-f_c^2}}\bigg[
2(1-f_c^2)\left(H_{3/2}^{(1)} -x_* H_{1/2}^{(1)} \right)H_{-\nu_\sigma}^{(2)}\, \\
&\qquad\qquad\qquad+e^{-\frac{i}{2}\pi \nu_c}\left((1+2f_c^2-2\nu_\sigma)H_{-\nu_\sigma}^{(2)}+2x_* H_{1-\nu_\sigma}^{(2)}\right) H^{(1)}_{\nu_c}
\bigg],
\\
\bar \delta^*_k&=\frac{-i\pi\, e^{i \pi \nu_\sigma}}{8\sqrt{1-f_c^2}}\bigg[
2(1-f_c^2)\left(H_{3/2}^{(1)} -x_* H_{1/2}^{(1)} \right)H_{\nu_\sigma}^{(1)} \\
&\qquad\qquad\qquad+e^{-\frac{i}{2}\pi \nu_c}\left((1+2f_c^2+2\nu_\sigma)H_{\nu_\sigma}^{(1)}+2x_* H_{1+\nu_\sigma}^{(1)}\right) H^{(1)}_{\nu_c}
\bigg].
\end{align}
It can be shown that in both cases, these coefficients satisfy eqs.~\eqref{cond1}-\eqref{cond4}.

With this set of coefficients we see that $\mathcal{C}^{\phi \phi}_{k}$, $\mathcal{C}^{\sigma \sigma}_{k}$ are different than those in eqs.~\eqref{Cabh}, and therefore the initial conditions of the entangled state produced by this simple model is different from those adopted in the ABH work while both have entanglement of the same type (\ref{eqn:entangled}).
On the other hand, from the general analysis in the previous section, it is conceivable that there should exist a family of perhaps more contrived models that result in a state that is closer to the ABH state for a range of $k$.    As shown in (\ref{Cabh}) and Appendix~\ref{app:Bogcoeff}, one can at least find a set of generalized Bogoliubov coefficients that satisfy the relations (\ref{cond1})-(\ref{cond4}) and that results in the exact ABH state.  When working with these types of Gaussian states, or equivalently with generalized Bogoliubov transformations, we are operating at the level of the quadratic action, which for a two-field model is in general specified by a $2\times 2$ symmetric kinetic matrix, a $2\times 2$ anti-symmetric friction matrix and a $2\times 2$ symmetric squared mass matrix in the Fourier space. All components of the three matrices are time-dependent in general and thus the quadratic action includes $7$ independent functions of time for each $k$. Since a full multi-field effective action (possibly after integrating out other heavy fields) incorporates all orders of the fields and derivatives, expanding the full action around a homogeneous and isotropic but time-dependent background could end up with a rather non-trivial quadratic action in general. While a system  described by a Gaussian state (or  generalized  Bogoliubov transformations) is specified by a finite set of functions of $k$, a whole multi-field action has more degrees of freedom, in particular it is specified by an infinite set of functions of the fields.  It is thus interesting to ask whether there exists a  multi-field model that exactly or approximately results in a  given set of generalized Bogoliubov coefficients for a range of $k$ which reproduce the ABH state, by trading off functions of the fields (i.e. terms in the multi-field Lagrangian) with functions of $k$ (i.e. generalized Bogoliubov coefficients). One of the main messages of this section thus far is that entanglement contained in the ABH state and that naturally emerges from the multi-field dynamics are of the same type shown in (\ref{eqn:entangled}).

Although in this section we calculate in the Heisenberg picture,
one can obtain the same result through the Schr\"odinger picture.
The time evolution of the entangled state, for $\eta\geq \eta_0$, is governed by the Schr\"{o}dinger equation with the free Hamiltonians of the two fields $\phi$ and $\sigma$,
while,  for $\eta < \eta_0$, the kinetic mixing between them modifies the evolution of the state from the case with their free Hamiltonians.
%Although there is no direct coupling in the Hamiltonian, at this time, the entanglement in the state  produces small oscillations in the power spectrum
For a further look at how entanglement is dynamically induced by kinetic mixing in the action, from the perspective of the Schr\"{o}dinger picture, see Appendix~\ref{app:Schrodinger}.

%====================================================================================%
\subsection{Power spectrum of inflaton perturbations}
%====================================================================================%

\begin{figure}[tbp]
  \begin{center}
  \includegraphics[width=100mm]{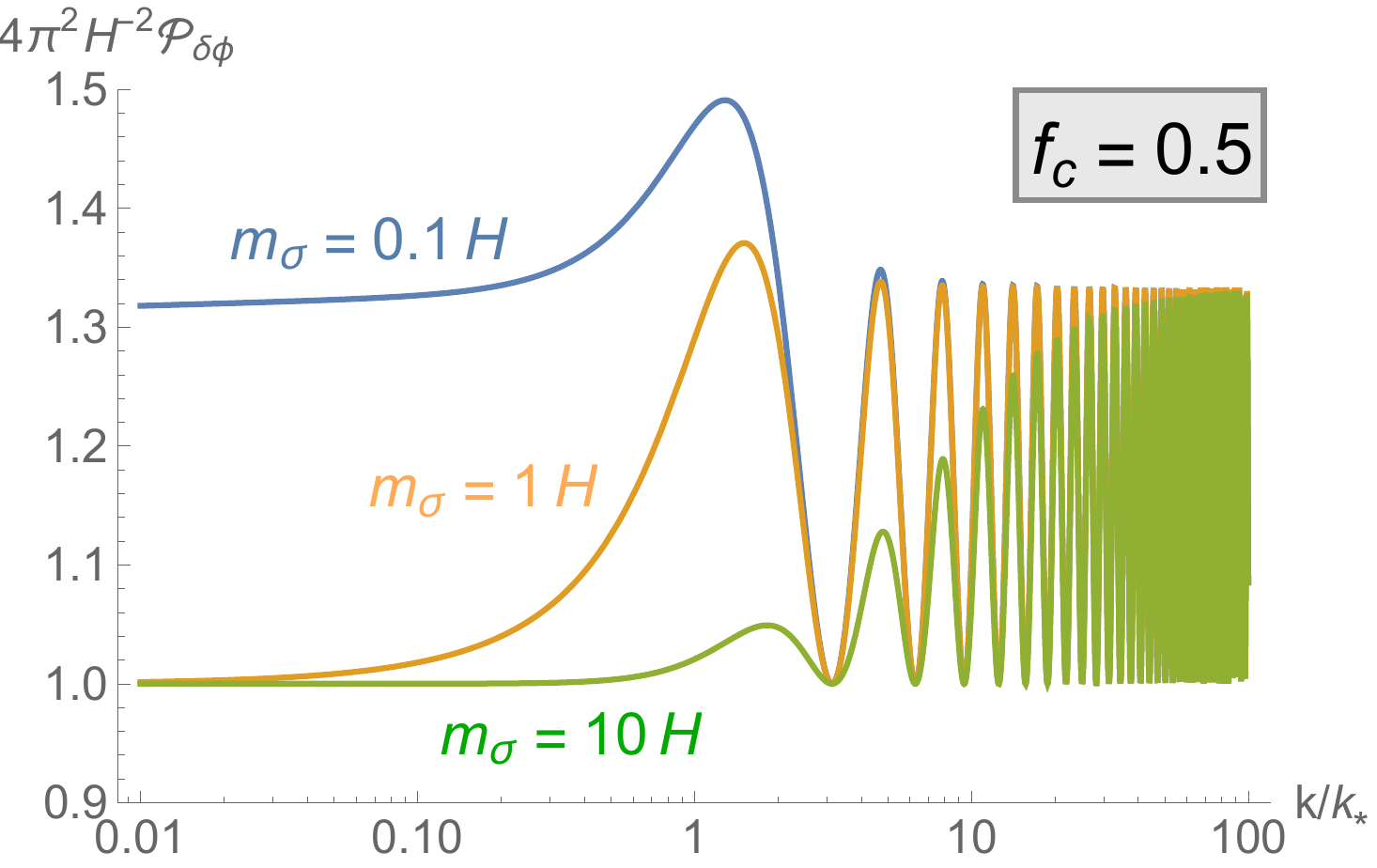}
  \end{center}
  \caption
 {The dimensionless power spectrum of the massless scalar field $\delta\phi$ normalized by $(H/2\pi)^2$. We set $f_c=0.5$ and $m_\sigma=0.1H$ (blue),
$1H$ (yellow) and $10H$ (green). The horizontal axis is $k/k_*$ where $k_*$ corresponds to a mode 
that exits the horizon when $f(\eta)$ vanishes.}
 \label{Power}
\end{figure}

Here, we obtain the power spectrum of the inflaton perturbations, which is regarded to be
connected with the curvature perturbations observed by the Cosmic Microwave Background radiation.
Now that we have obtained the expressions for the coefficients of  $\hat{a}_{\bm k} ^{(\Phi)}$
and $\hat{b}_{\bm k} ^{(S)}$ in eqs.~\eqref{A expression} and \eqref{B expression}
in this model, it is simpler to calculate based on the two-point function of $\hat{u}_{\phi}(\eta,\bm k)$
with respect to the in-vacuum state directly, although we can also calculate based on 
eqref.~\ref{inoutvac} and the creation/annihilation operators of $\phi$ and $\sigma$.
%the reduced density matrix obtained in subsection~\ref{subsec reduced density matrix}.
The two-point function of $\hat{u}_{\phi}(\eta,\bm k)$ is
\begin{align}
{}_{\rm in}\langle 0|\hat{u}_{\phi}(\eta,\bm k) \hat{u}_{\phi}(\eta,\bm k')
|0\rangle_{\rm in}=(2\pi)^3\delta(\bm k+\bm k')\times
\biggl[|\alpha_k u_{\phi k}+\beta^* _k  u_{\phi k}^*|^2
+|\gamma_k u_{\phi k}+\delta^* _k u_{\phi k}^*|^2 \biggr],
\end{align}
where the first term in parentheses denotes the contribution from the one-particle state of $\Phi$,
while the second term denotes the contribution from that of $S$.
Since $u_\phi = a \delta \phi$ and $u_{\phi k} \;(x\to 0)= -i/(\sqrt{2k}x)$,
the power spectrum of $\delta\phi$ in the super-horizon limit is given by
\begin{equation}
\lim_{x\to0}\mcP_{\delta\phi}(k)=\left(\frac{H}{2\pi}\right)^2
\left(|\alpha_k-\beta^* _k |^2+|\gamma_k-\delta_k^*|^2\right).
\end{equation}
In figure~\ref{Power}, we plot the dimensionless power spectrum 
$(H/2\pi)^{-2}\mcP_{\delta\phi}$ in the current model.
An oscillatory feature is produced for the modes which are inside  the horizon when the fields are mixed at $\eta=\eta_*$.\footnote{In this simple model, since we assume that $f(\eta)$ instantaneously vanishes,
  modes with infinitely high-$k$ are excited.} Nevertheless, we expect that a UV cut-off scale
corresponding to the time scale of the $f(\eta)$ transition would appear in a more realistic model.
%Note that since $\beta_k=0$ in our model, the deviation from $\mcP_{\delta\phi}=(H/2\pi)^2$
%comes mostly from the effect of entanglement.
This result suggests that the oscillations in the power spectrum of the inflaton perturbation
reported in ABH \cite{Albrecht:2014aga} is qualitatively quite generic to initial states with
quantum entanglement between the inflaton and another scalar field.
%
%///////////////////////////////////////////////////////////////////////////////////%
%///////////////////////////////////////////////////////////////////////////////////%
%

%====================================================================================%
\section{Conclusions and Discussions \label{sec_conclusion}}
%====================================================================================%

In fundamental theories like supergravity or string theory, scalar fields are ubiquitous and
in some cases, some fields can affect each other through entanglement, even if 
they are decoupled at the level of the action.  In this work, we have studied the cosmological consequences
of  an entangled initial state between two fields $\phi$ (inflaton) and $\sigma$ (spectator)
in inflation. The perturbations of $\phi$  contribute to the curvature perturbation, 
while multi-field dynamics with the spectator field $\sigma$ affect observables through particle production
and entanglement. We use the general action in eq.~(\ref{first action}) together with the assumption that
the time variations of $G_{IJ} (t)$ (the field space metric) and $M_{IJ} (t)$ (the mass matrix)
are sufficiently small at early and late times. Under this assumption, we can diagonalize
 $G_{IJ} (t)$ and $M_{IJ} (t)$ both at late times with a set of fields $\{\phi, \sigma\}$, and early times with another set of fields
 $\{\Phi, S \}$. We then construct the out-vacuum defined in eq.~(\ref{out vac def}) and the in-vacuum defined in eq.~(\ref{def_in_vacuum})(on sufficiently small scales). 
 
With this setup, we have developed a formalism which defines the quantum entanglement
between the perturbations of $\phi$ and $\sigma$. The starting point of our formalism
is eqs.~(\ref{A expression}) and (\ref{B expression}) where
the annihilation operators for the out-vacuum are written as the linear combination of all the creation/annihilation operators for the in-vacuum.
Since we can compute the constant coefficients  
$\alpha_k, \beta_k, \gamma_k,\delta_k, \bar\alpha_k, \bar\beta_k, \bar\gamma_k,$ and $\bar\delta_k$
for any time evolution of $G_{IJ}$ and $M_{IJ}$, our formalism can be applied to a very general class of models.

In order to clarify the meaning of  eqs.~(\ref{A expression}) and (\ref{B expression}), 
we have expressed the in-vacuum state as an excited state of the out-vacuum state in eq.~(\ref{inoutvac})
with $f_{\bm k}$ given by eq.~(\ref{expression fk}) and  
$ \mathcal{C}^{\phi \phi}_{k}$, $\mathcal{C}^{\sigma \sigma}_{k}$ 
and $\mathcal{C}^{\phi \sigma}_{k}$ given by eq.~(\ref{C parameters}).
While the terms in eq.~(\ref{expression fk}) with $ \mathcal{C}^{\phi \phi}_{k}$
and $\mathcal{C}^{\sigma \sigma}_{k}$ lead to a vacuum squeezed state and can be understood
as a generalization of the Bogoliubov transformation, 
the term with  $\mathcal{C}^{\phi \sigma}_{k}$ produces cross terms proportional to
$(\hat{a}_{-\bm {k}} ^{(\phi)} )^\dag (\hat{b}_{\bm {k}} ^{(\sigma)} )^\dag+
(\hat{a}_{\bm {k}} ^{(\phi)} )^\dag (\hat{b}_{-\bm {k}} ^{(\sigma)} )^\dag $ which result in an entangled state. 
Since $ \mathcal{C}^{\phi \phi}_{k}$, $\mathcal{C}^{\sigma \sigma}_{k}$ 
and $\mathcal{C}^{\phi \sigma}_{k}$ are expressed in terms of 
$\alpha_k, \beta_k, \gamma_k,\delta_k, \bar\alpha_k, \bar\beta_k, \bar\gamma_k,$ and $\bar\delta_k$,
we have shown that as long as the two sets of creation/annihilation operators
are related by eqs.~(\ref{A expression}) and (\ref{B expression}), in general,
multi-field dynamics induce particle production of $\phi_{\bf k}$, $\sigma_{\bf k}$ particles
as well as quantum entanglement between them. In particular, we have concluded that
the quantum fluctuations of $\phi$ and $\sigma$ are entangled if the condition 
(\ref{entangle condition}) is satisfied. 

Although the above result is obtained in the Heisenberg picture, of course,
we have been able to relate our conclusion to the Schr\"odinger picture result
by constructing the wave functions of the in-vacuum and out-vacuum.
Then, we compared our in-vacuum wave function with the entangled state
considered in the previous work \cite{Albrecht:2014aga} (ABH).
We have shown that, at $\eta=\eta_0$,  our in-vacuum wave function coincides with  the entangled Gaussian state adopted in ABH if 
 the constants $ \mathcal{C}^{\phi \phi}_{k}$,  $\mathcal{C}^{\sigma \sigma}_{k}$ and $\mathcal{C}^{\phi \sigma}_{k}$ are fixed to eqs.~\eqref{Cabh}. Having done so, the initial conditions of both the state anzatz used in ABH and the one described in this work are the same, and hence, the dynamics of both states will be the same when evolved with the Schr\"{o}dinger equation for $\eta \geq \eta_0$   with the free Hamiltonian.

Finally, we considered a concrete example with a sudden  change
in the kinetic matrix of the scalar fields given by eqs.~(\ref{example GIJ}) and (\ref{example ft}).
From the junction condition, which requires that the mode functions and their derivatives
are continuous at the time when the off-diagonal components disappear,
we obtained the constants
$\alpha_k, \beta_k, \gamma_k,\delta_k, \bar\alpha_k, \bar\beta_k, \bar\gamma_k,$ and $\bar\delta_k$.
We have shown that in this example, $\phi$ and $\sigma$ are entangled in the in-vacuum state.
Then we calculated the power spectrum of the inflaton
and confirmed that  an oscillatory feature appears, akin to that presented in ref.~\cite{Albrecht:2014aga}.

Since we are mainly interested in establishing a general argument on quantum entanglement  in multi-field inflation, for simplicity, in the concrete example we introduced the time dependence of 
$G_{IJ} (t)$ by hand. We also considered the perturbation of scalar fields in a fixed de Sitter background and assumed that
the fluctuations of the inflaton field are related with the observable curvature perturbation,
following ref.~\cite{Albrecht:2014aga}.  We expect that there are more realistic models where the time evolution of $G_{IJ}$
arises from dynamics of scalar fields other than $\phi$ and $\sigma$.  It would be interesting to investigate whether the consistent model can be obtained from a more fundamental theory.

We have shown that oscillatory features in the power spectrum of the inflaton perturbation
are a general prediction of an initial entangled state. In their turn, these oscillations translate into oscillations in the angular power spectrum \cite{Bolis:2016vas} which can be used to compare directly to CMB data 
\cite{Ade:2015xua,Ade:2015ava}, however, it is well known that similar oscillatory features can also be produced in multi-field inflation models with a sudden turn
 \cite{Tolley:2009fg,Cremonini:2010ua,Achucarro:2010da,Shiu:2011qw,Pi:2012gf,Gao:2012uq,Noumi:2012vr}.
It is therefore important to consider how to distinguish 
and/or understand degeneracies between these two sets of models by using the formula (\ref{Cabh}) and the correspondence summarized in Appendix~\ref{app:Bogcoeff}. In particular, it is interesting to ask whether there exists a multi-field model that exactly or approximately results in the same quadratic order behavior given by the ABH state (or similar entangled states).

%\textcolor{red}{a given set of generalized Bogoliubov coefficients for a range of $k$ by trading off functions of the fields (i.e. terms in the multi-field Lagrangian) with functions of $k$ (i.e. generalized Bogoliubov coefficients).}

One possible way to phenomenologically  distinguish between these models is to look at the non-Gaussianity signals produced by entangled states and multi-field inflation models.
In single-field inflation, the effect of a deviation
from the Bunch Davies vacuum produces relatively large primordial non-Gaussianity
\cite{Chen:2006nt,Holman:2007na, Agullo:2010ws,Ashoorioon:2010xg,Ashoorioon:2011eg,Agarwal:2012mq}. It may also be possible that
the effect of an entangled initial state produces large non-Gaussianity that can be used to constrain these models with our current and upcoming CMB data~\cite{Meerburg:2009ys,Ganc:2011dy}.
More importantly, it is intriguing to ask whether entanglement can produce a different characteristic shape of non-Gaussianity from multi-field inflation alone, and therefore such a signal could be used to distinguish the two scenarios (see upcoming paper by ABH). Since we have shown that an entangled state can naturally emerge from multi-field dynamics, it is conceivable to expect that models of multi-field inflation can generate the type of non-Gaussianities that an entangled initial state predicts. (If the late-time setups for the two models have the same Lagrangian and the same quantum state then all observables including those associated with non-Gaussianities are obviously the same.) While it is easy to parameterize the entangled initial state e.g. as in (\ref{eqn:entangled}) as a phenomenological description at late time, it is useful to relate the multi-field models to more fundamental theories such as string theory. In this sense the two sets of models, the entangled state and the multi-field inflation, are complementary.
This will also become an important tool to access information of nonlinearities in the hidden sector, like \cite{Assassi:2013gxa}. We would like to leave these topics for future works.

%************************* Acknowledgments *****************************%
\acknowledgments
We thank Andreas Albrecht, Rich Holman, Keisuke Izumi, Sugumi Kanno, Yasusada Nambu, Jiro Soda, Takahiro Tanaka, and Kazuhiro Yamamoto,  for valuable comments and useful suggestions. NB acknowledges funding from the European Research Council under the European Union's Seventh Framework Programme (FP7/2007-2013)/ERC Grant Agreement No. 617656 ``Theories and Models of the Dark Sector: Dark Matter, Dark Energy and Gravity''.
The work of T. F. was supported by Japan Society for the Promotion of Science (JSPS) Grants-in-Aid for Scientific Research (KAKENHI) No. 17J09103.
The work of S. Mizuno was supported by Japan Society for the Promotion of Science (JSPS) Grants-in-Aid for Scientific Research (KAKENHI) No. 16K17709. The work of S. Mukohyama was supported by Japan Society for the Promotion of Science (JSPS) Grants-in-Aid for Scientific Research (KAKENHI) No. 17H02890, No. 17H06359, and by World Premier International Research Center Initiative (WPI), MEXT, Japan.

\appendix

\section{Generalized Bogoliubov coefficients in terms of ($C_k^{\phi\phi}$, $C_k^{\sigma\sigma}$, $C_k^{\phi\sigma}$)}
\label{app:Bogcoeff}

For the trivial values of the generalized Bogoliubov coefficients ($\alpha_k=1$, $\beta_k=0$, $\gamma_k=0$, $\delta_k=0$, $\bar{\alpha}_k=0$, $\bar{\beta}_k=0$, $\bar{\gamma}_k=1$, $\bar{\delta}_k=0$), the relations (\ref{cond1})-(\ref{cond4}) are satisfied and ($C_k^{\phi\phi}$, $C_k^{\sigma\sigma}$, $C_k^{\phi\sigma}$) defined by (\ref{C parameters}) vanish.

Let us now expand the generalized Bogoliubov coefficients and ($C_k^{\phi\phi}$, $C_k^{\sigma\sigma}$, $C_k^{\phi\sigma}$) around the trivial values as
\begin{eqnarray}
& \alpha_k = 1 + \epsilon \alpha_{k,1} + \epsilon^2 \alpha_{k,2} + \mathcal{O}(\epsilon^3)\,, \quad 
  \beta_k = \epsilon \beta_{k,1} + \epsilon^2 \beta_{k,2} + \mathcal{O}(\epsilon^3)\,,\nonumber\\
& \gamma_k = \epsilon \gamma_{k,1} + \epsilon^2 \gamma_{k,2} + \mathcal{O}(\epsilon^3)\,,\quad
  \delta_k = \epsilon \delta_{k,1} + \epsilon^2 \delta_{k,2} + \mathcal{O}(\epsilon^3)\,,\quad\nonumber\\
& \bar{\alpha}_k = \epsilon \bar{\alpha}_{k,1} + \epsilon^2 \bar{\alpha}_{k,2} + \mathcal{O}(\epsilon^3)\,, \quad 
  \bar{\beta}_k = \epsilon \bar{\beta}_{k,1} + \epsilon^2 \bar{\beta}_{k,2} + \mathcal{O}(\epsilon^3)\,,\nonumber\\
 & \bar{\gamma}_k = 1 + \epsilon \bar{\gamma}_{k,1} + \epsilon^2 \bar{\gamma}_{k,2} + \mathcal{O}(\epsilon^3)\,,\quad
  \bar{\delta}_k = \epsilon \bar{\delta}_{k,1} + \epsilon^2 \bar{\delta}_{k,2} + \mathcal{O}(\epsilon^3)\,, \quad
\end{eqnarray}
and
\begin{equation}
 C_k^{\phi\phi} =  \epsilon C_{k,1}^{\phi\phi} + \epsilon^2 C_{k,2}^{\phi\phi} + \mathcal{O}(\epsilon^3)\,, \quad
  C_k^{\sigma\sigma} =  \epsilon C_{k,1}^{\sigma\sigma} + \epsilon^2 C_{k,2}^{\sigma\sigma} + \mathcal{O}(\epsilon^3)\,, \quad
    C_k^{\phi\sigma} =  \epsilon C_{k,1}^{\phi\sigma} + \epsilon^2 C_{k,2}^{\phi\sigma} + \mathcal{O}(\epsilon^3)\,, 
\end{equation}
where $\epsilon$ is a small bookkeeping parameter. It is then easy to solve (\ref{cond1})-(\ref{cond4}) and (\ref{C parameters}) up to $\mathcal{O}(\epsilon)$ as
\begin{eqnarray}
 & \Re \alpha_{k,1} = 0\,,\quad
  \beta_{k,1} = C_{k,1}^{\phi\phi}\,,\quad
  \delta_{k,1}  = C_{k,1}^{\phi\sigma}\,,\nonumber\\
 & \bar{\alpha}_{k,1} = -\gamma_{k,1}^*\,,\quad
  \bar{\beta}_{k,1} = C_{k,1}^{\phi\sigma}\,,\quad
  \Re \bar{\gamma}_{k,1} = 0\,, \quad
  \bar{\delta}_{k,1}  = C_{k,1}^{\sigma\sigma}\,, 
\end{eqnarray}
where $\Im\alpha_{k,1}$, $\gamma_{k,1}$ and $\Im\bar{\gamma}_{k,1}$ are freely specifiable, and a superscript $*$ represents complex conjugate. This means that there exists a family of generalized Bogoliubov coefficients that corresponds to ($C_k^{\phi\phi}$, $C_k^{\sigma\sigma}$, $C_k^{\phi\sigma}$) up to order $\mathcal{O}(\epsilon)$.

It is also straightforward to solve (\ref{cond1})-(\ref{cond4}) and (\ref{C parameters}) up to $\mathcal{O}(\epsilon^2)$. For $\Im\alpha_{k,1} = \gamma_{k,1} = \Im\bar{\gamma}_{k,1} = 0$, the solution is
\begin{eqnarray}
 & \Re \alpha_{k,2} = \frac{1}{2}\left(\left|C_{k,1}^{\phi\phi}\right|^2+\left|C_{k,1}^{\phi\sigma}\right|^2\right)\,,\quad
  \beta_{k,2} = C_{k,2}^{\phi\phi}\,,\quad
  \delta_{k,1}  = C_{k,2}^{\phi\sigma}\,,\nonumber\\
 & \bar{\alpha}_{k,2} = C_{k,2}^{\phi\phi *}C_{k,2}^{\phi\sigma}+C_{k,2}^{\phi\sigma *}C_{k,2}^{\sigma\sigma} -\gamma_{k,2}^*\,,\quad
 \bar{\beta}_{k,2} = C_{k,2}^{\phi\sigma}\,,\nonumber\\
 & \Re \bar{\gamma}_{k,1} = \frac{1}{2}\left(\left|C_{k,1}^{\phi\sigma}\right|^2+\left|C_{k,1}^{\sigma\sigma}\right|^2\right)\,,\quad
  \bar{\delta}_{k,2}  = C_{k,2}^{\sigma\sigma}\,, 
\end{eqnarray}
where $\Im\alpha_{k,2}$, $\gamma_{k,2}$ and $\Im\bar{\gamma}_{k,2}$ are freely specifiable. This means that there exists a family of generalized Bogoliubov coefficients that corresponds to ($C_k^{\phi\phi}$, $C_k^{\sigma\sigma}$, $C_k^{\phi\sigma}$) up to order $\mathcal{O}(\epsilon^2)$. Obviously, one can continue the same procedure up to any order in $\epsilon$.

 \section{Entanglement induced from kinetic mixing: Schr\"{o}dinger perspective.}
\label{app:Schrodinger}
To further illustrate how kinetic mixing can induce an entangled state, even once the coupling in the action has vanished, we consider a similar model to that
%general scenario 
discussed in Sec.\ref{sec_concrete model}  from the perspective of the Schr\"{o}dinger picture.   Again, the final state that evolves according to the whole Hamiltonian will be the initial state for the evolution with the free Hamiltonian.

Here we consider the following example with kinetic mixing in the action of eq.~\eqref{first action} given by:
\begin{equation}
G_{IJ}(\eta)=\begin{pmatrix}1 & f(
\eta) \\
f(\eta) & 1 \\
\end{pmatrix},
\qquad
M_{IJ}=\begin{pmatrix}m_{\phi}^2 & 0 \\
0 & m_\sigma^2 \\
\end{pmatrix}.
\end{equation}
As described above, for all times $\eta\geq \eta_0$ we set $f(\eta)=0$ allowing $G_{IJ}$ to be diagonal. However, for $\eta<\eta_0$ the whole action includes $f(\eta)\neq 0$ and therefore there is kinetic mixing. To define the Hamiltonian for this system we first need to find the conjugate momenta for the fields:
\begin{eqnarray}
\Pi_{\phi_k} = \frac{\partial \mathcal{L}}{\partial \phi_{-k}} = a^2(\phi^{\prime}_k + f(t)\sigma_k^{\prime})\\
\Pi_{\sigma_k} = \frac{\partial \mathcal{L}}{\partial \sigma_{-k}} = a^2(\sigma^{\prime}_k + f(t)\phi_k^{\prime})
\end{eqnarray}
%Invert them to get $\sigma^{\prime}_k ,\phi_k^{\prime}$:
%\begin{eqnarray}
%\phi_k^{\prime}= \frac{\Pi_{\phi_k}- f(t)\Pi_{\sigma_k} }{a^2(1-f(t)^2)}\\
%\sigma_k^{\prime}= \frac{\Pi_{\sigma_k}- f(t)\Pi_{\phi_k} }{a^2(1-f(t)^2)}
%\end{eqnarray}
The Hamiltonian then takes the form:
\begin{eqnarray}
H_{\bk} &=& \frac{1}{2} \frac{1}{a^2 (1-f(\eta)^2)} [\Pi_{\phi_{\bk}}\Pi_{\phi_{-\bk}}+\Pi_{\sigma_{\bk}}\Pi_{\sigma_{-\bk}}-f(\eta)(\Pi_{\phi_{\bk}}\Pi_{\sigma_{-\bk}}+\Pi_{\sigma_{\bk}}\Pi_{\phi_{-\bk}})] \nonumber\\
&+& \frac{a^2}{2} [\omega_{\phi}^2 \phi_{\bk} \phi_{-\bk} + \omega_{\sigma}^2 \sigma_{\bk} \sigma_{-\bk} +f(\eta) \omega_{\phi\sigma}^2 (\phi_{\bk}\sigma_{-\bk}+\sigma_{\bk} \phi_{-\bk})]
\end{eqnarray}
where $\omega_{\phi}^2 =  k^2 + a^2 m_{\phi}^2, \;\omega_{\sigma}^2 =  k^2 + a^2 m_{\sigma}^2$, and $\omega_{\phi \sigma}^2 =  k^2 $.
Since there is kinetic mixing in the action for $\eta<\eta_0$, the Hamiltonian also has mixed conjugate momentum terms.  The corresponding wavefunction that will solve the Schr\"{o}dinger equation with this Hamiltonian will then also need a cross term between the two fields. Such a state takes the form,
\begin{equation}
\Psi_k = \mathcal{N}_{\bk} \exp\left[-\frac{1}{2}\left( A_k(\eta) \phi_{\bk} \phi_{-\bk}  + B_k(\eta) \sigma_{\bk} \sigma_{-\bk} +  C_k(\eta) (\phi_{\bk} \sigma_{-\bk}+ \phi_{-\bk} \sigma_{\bk}) \right)\right]
\end{equation}
Finally, plugging this state into the Schr\"{o}dinger equation:
\begin{equation}
i \partial_{\eta} \Psi_{\bk}(\phi, \sigma ;\eta) = H_{\bk} \Psi_{\bk}(\phi, \sigma ;\eta)
\end{equation}
results in the following equations of motion for the state parameters at $\eta<\eta_0$:
\begin{eqnarray}
i A_k^{\prime} &=& \frac{A_k^2+C_k^2}{a^2 (1-f(\eta)^2)} - 2 f(\eta) \frac{A_k C_k}{a^2(1-f(\eta)^2)} - a^2 \omega_{\phi}^2 \label{Aeqt}\\
i B_k^{\prime} &=& \frac{B_k^2+C_k^2}{a^2 (1-f(\eta)^2)} - 2 f(\eta) \frac{B_k C_k}{a^2(1-f(\eta)^2)} - a^2 \omega_{\sigma}^2 \label{Beqt}\\
i C_k^{\prime} &=& C_k\frac{A_k+B_k}{a^2 (1-f(\eta)^2)} -  f(\eta)\frac{A_k B_k+C_k^2}{a^2(1-f(\eta)^2)} - f(\eta) a^2 \omega_{\phi\sigma}^2 \label{Ceqt}
\end{eqnarray}

Looking at these equations one can get an ulterior perspective on how entanglement is generated in such a system. If both $f(\eta)$ and $C_k(\eta)$ are zero there would be no coupling in the Hamiltonian or entanglement in the state, and one would recover the standard Bunch-Davies solution. On the other hand, if the coupling $f(\eta)$ in the Hamiltonian is nonzero, but the entanglement of the state is set to zero initially (i.e. $C_k(\eta_{\text{init}}) = 0$), one can see from the coupled equations of motion, that the coupling $f(\eta)$ induces the mixing term proportional to $C_k(\eta)$ in the state to evolve and take on nonzero values. Then, when $f(t)$ vanishes at $\eta=\eta_0$, the Hamiltonian will no longer be coupled, however, the mixing (or entanglement) term $C_k(\eta)$ in the state now has a non zero value, and will perpetuate the entanglement between the two fields. Note that the solutions of the state coefficients $A_k, B_k$ and $C_k$ that solve the above equations for $\eta<\eta_0$ are different than those defined for $\eta\geq \eta_0$ when the Hamiltonian of the two fields is decoupled. 

This is consistent with the results we obtained in Section \ref{sec_schrodinger} where we determined that for late times, when the Hamiltonian of $\phi$ and $\sigma$ is decoupled, the state of  these fields is still entangled. Note that in general, the initial conditions for the state parameters $A_k, B_k$ and $C_k$ at $\eta=\eta_0$ will not be the same as those used in ABH or equivalently those in eqs.~\eqref{init_condit}. The values of $A_k, B_k$ and $C_k$ here evolve from the dynamics induced by the full Hamiltonian with the coupling before $\eta_0$. This model is clearly a more general scenario then the one explored in ABH, however it does help show what kind of mechanisms may give rise to entangled states of that form.

%%%%%%%%%%%%%%%%%%%%%%%%%%%%%%%%%%%%%%%%%%%%%%%%%%%%%%%%%%%%%%%%%%%%%%%%%%%%%%%%%%%%%%%%%%%%%%%%%%%

%%%%%%%%%%%%%%%%%%%%%%%%%%%%%%%%%%%%%%%%%%%%%%%%%%%%%%%%%%%%%%%%%%%%%%%%%%%%%%%%%%%%%%%%%%%%%%%%%%%

\end{document}